\documentclass[aps,prb,twocolumn,showpacs,floatfix]{revtex4-1}
\usepackage{amsmath,amsfonts,amssymb}
\usepackage[english]{babel}

\usepackage{graphicx,color,fancyhdr,xspace}
\definecolor{darkgreen}{rgb}{0,0.5,0}
\definecolor{darkblue}{rgb}{0,0,0.2}

\RequirePackage[
   hyperindex,colorlinks,bookmarksnumbered,
   plainpages=true,pdfstartview=FitH]{hyperref}
\hypersetup{linkcolor=blue,urlcolor=darkblue,citecolor=darkgreen}
\usepackage[pdfstartview=FitH]{hyperref}

   \def\wrt{\emph{w.r.t.}\@\xspace}
   \def\eg{\emph{e.\,g.}\@\xspace}
   \def\ie{\emph{i.\,e.}\@\xspace}
   \def\cf{\emph{cf.}\@\xspace}
   \def\vs{\emph{vs.}\@\xspace}

   \def\Jp{\ensuremath{J^{\prime}}\xspace}
   \def\Jpc{\ensuremath{J^{\prime}_{^{_{\mathrm{C}}}}}\xspace}

   \def\qJ{\ensuremath{q_J}\xspace}
   \def\qJrel{\ensuremath{\tilde{q}_J}\xspace}
   \def\qJclass{\ensuremath{q^{\mathit{cl}}_J}\xspace}
   \def\Eclass{\ensuremath{E_q^{\mathit{cl}}}\xspace}
   \def\E0class{\ensuremath{E_0^{\mathit{cl}}}\xspace}

   \def\CuCl{\ensuremath{\mathrm{Cs_{2} Cu Cl_{4}}}}
   \def\kETorganic{\ensuremath{\mathrm{\kappa-(ET)_2 Cu_2 (CN)_3}}\xspace}

   \def\Szx{\ensuremath{\langle S_{z,x}\rangle}\xspace}
   \def\Deff{\ensuremath{D^{\ast}}\xspace}

   \def\nc{\ensuremath{n^{_{\mathrm{C}}}_{^{_\mathrm{S}}}}\xspace}
   \def\Ly{\ensuremath{n_{^{_\mathrm{C}}}}\xspace}

   \def\x{\zeta}
   \def\cpsBC{cps-BC\xspace}  
   \def\perBC{per-BC\xspace}  

   \renewcommand{\exp}[1]{e^{#1}}

   \def\SS{\ensuremath{\langle S\cdot S\rangle}\xspace}
   \def\SSJp{\ensuremath{\langle S\cdot S\rangle_{J^{\prime}}}\xspace}
   \def\SSJ{\ensuremath{\langle S\cdot S\rangle_{J}}\xspace}
   \def\SSJpa{\ensuremath{\langle\langle S\cdot S\rangle_{J^{\prime}}\rangle}\xspace}
   \def\SSJa{\ensuremath{\langle\langle S\cdot S\rangle_{J}\rangle}\xspace}

   \newcommand{\Eq}[1]{Eq.~(\ref{#1})\xspace}

   \newcommand{\Fig}[1]{Fig.~\ref{#1}}
   \newcommand{\Figp}[2]{Fig.~\ref{#1}(#2)}
   \newcommand{\Figs}[1]{Figs.~\ref{#1}}

\begin{document}

\title{
  Incommensurate correlations in the anisotropic triangular
  Heisenberg lattice
}

\author{Andreas \surname{Weichselbaum}}
\affiliation{
  Physics Department, Arnold Sommerfeld Center for Theoretical Physics, and
  Center for NanoScience, Ludwig-Maximilians-Universit\"at, 80333 Munich,
  Germany
}
\author{Steven R. \surname{White}}
\affiliation{
  Department of Physics and Astronomy, University of California,
  Irvine, CA 92697, USA
}

\begin{abstract}
We study the anisotropic spin-half antiferromagnetic triangular 
Heisenberg lattice in two dimensions, seen as a set of chains with 
couplings $J$ ($J^{\prime}$) along (in between) chains, respectively. 
Our focus is on the incommensurate correlation that emerges in this 
system in a wide parameter range due to the intrinsic frustration of 
the spins. We study this system with traditional DMRG using 
cylindrical boundary conditions to least constrain possible 
incommensurate order. Despite that the limit of essentially decoupled 
chains $J'/J \lesssim 0.5$ is not very accessible numerically, it 
appears that the spin-spin correlations remain incommensurate for any 
finite $0<J' < \Jpc$, where $\Jpc/J>1$. The incommensurate wave 
vector $\qJ$, however, approaches the commensurate value 
corresponding to the antiferromagnetic correlation of a single chain 
very rapidly with decreasing $J'/J$, roughly as $\qJ \sim \pi - c_1 
(J'/J)^n \exp{-c_2 J/J'}$. 
\end{abstract}


\pacs{
  75.10.Jm,  
  71.10.Pm,  
  75.40.Mg,  
  75.50.Ee   
}

\date{\today}
\maketitle

\section{Introduction}

\begin{figure}[b]
\begin{center}
\includegraphics[width=0.75\linewidth]{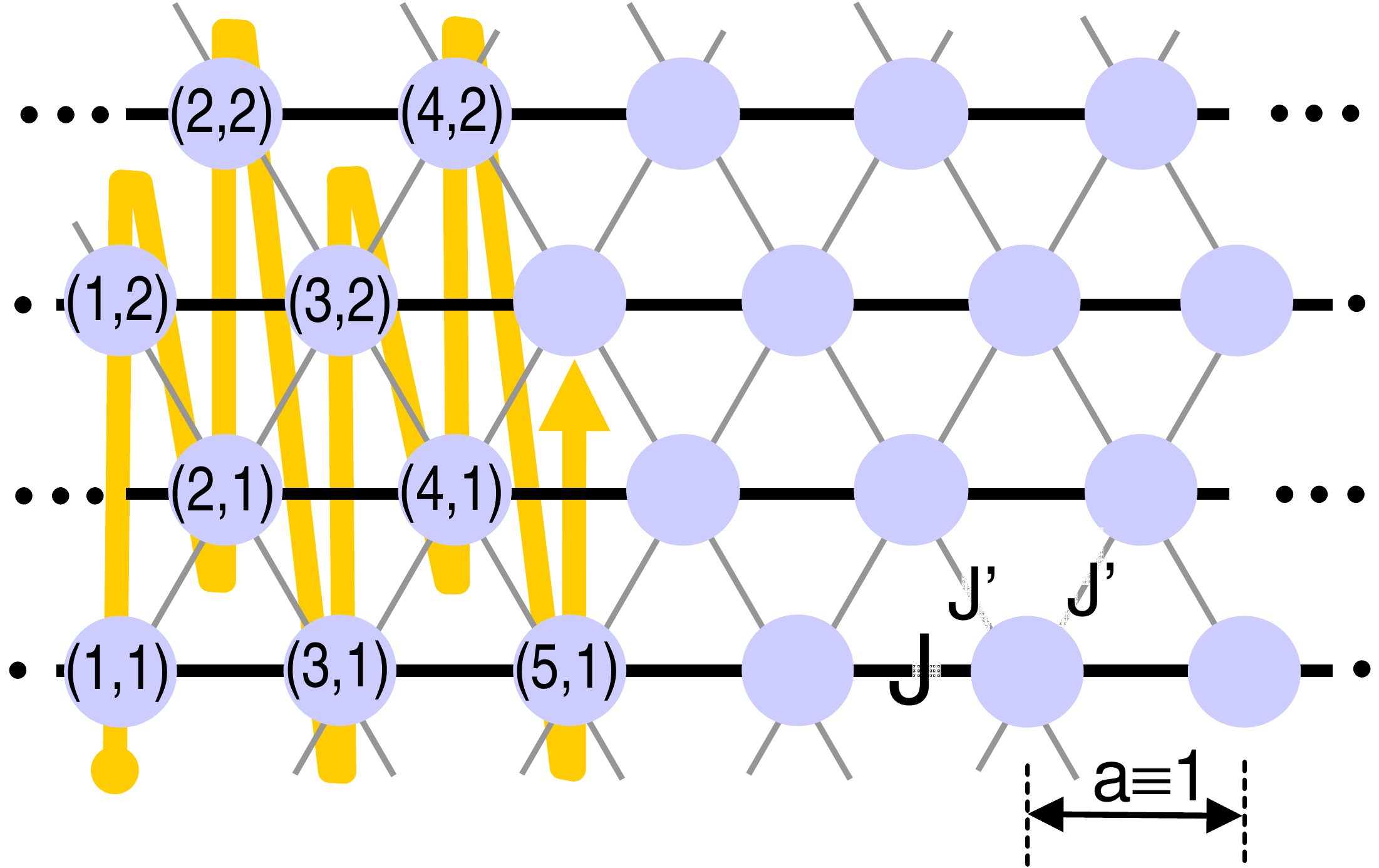}
\end{center}
\caption{
(Color online) The anisotropic triangular Heisenberg lattice
viewed as a set of parallel chains with intrachain coupling $J:=1$
and interchain coupling $J^{\prime}$ with lattice spacing $a := 1$.
For the numerical simulation using DMRG, cylindrical boundary
conditions with periodic wrapping in the transverse vertical
direction are assumed unless indicated otherwise. The
quasi-one-dimensional sweeping path through the triangular system
used within DMRG is indicated at the left side starting with site
$(1,1)$. This path is generalized to systems of different width.
}\label{fig:lattice} %
\end{figure}

The anisotropic triangular spin-1/2 Heisenberg lattice has been
suggested as an effective description for several organic and
anorganic compounds such as \CuCl \cite{Coldea02,Starykh10} or
\kETorganic. \cite{Shimizu03,Kandpal09,Qi09} These bulk systems
typically consist of layered structures with weak inter-layer
coupling, next-nearest neighbor and spin-orbit interactions. The
experimental observation of spin-liquid-like behavior in these
systems in certain parameter ranges therefore sparked renewed
interest in the anisotropic triangular model system. \cite{Balents10}
The simplest effective model is depicted schematically in
\Fig{fig:lattice}. It is viewed as a set of chains with intrachain
coupling $J$, that are coupled in planar triangular fashion by the
interchain coupling $J^{\prime}$. In the absence of an external
magnetic field, all energies can be written in units of $J:=1$, which
thus yields the single dimensionless coupling parameter $\Jp \equiv
J^{\prime}/J$, as used throughout this paper unless indicated
otherwise.
Extensive theoretical studies have been performed on this model
system, \cite{Weng06,Starykh07,Heidarian09} but the full phase
diagram has remained elusive, in particular for smaller \Jp.
Approximate numerical studies \cite{Heidarian09} found that the
magnetic order vanishes near $\Jp \lesssim 0.85$, with a possibly
continuous transition to an essentially one-dimensional collinear
phase for $\Jp \lesssim 0.6$ [\onlinecite{Heidarian09}] ($\Jp
\lesssim 0.3$, [\onlinecite{Ghamari11}]). The presence of collinear
versus incommensurate order at weak chain-coupling \Jp, thus remains
controversial, \cite{Weng06,Starykh07,Heidarian09,Ghamari11} and as
such represents a major motivation for this paper.

Here we present an extensive set of density matrix renormalization
group (DMRG) \cite{White92} calculations for ladders and cylinders
for this system with widths ranging from two to ten lattice spacings.
Recently the use of DMRG for frustrated two dimensional (2D) systems
has proven to be very powerful -- the results are highly precise and
unbiased for the narrower systems, and maintain acceptable accuracy
to widths of about ten or twelve chains. Careful consideration of
finite size effects have allowed strong conclusions about the 2D
ground state both in an antiferromagnetically ordered system (the
isotropic triangular Heisenberg model)\cite{White07} and for a spin
liquid (the kagome Heisenberg model). \cite{Yan11} For a review of
the techniques important for such 2D DMRG studies see
[\onlinecite{Stoudenmire11}]. Of course, each system is different,
and for the anisotropic triangular Heisenberg model we study here,
the incommensurate correlations and the associated finite size
effects must be dealt with carefully.

For that purpose, we chose as our primary type of cluster a
cylindrical geometry, with the cylinder's axis along the $J$
direction [\cf \Fig{fig:lattice}]. Despite our limitation to
relatively small circumferences, given the strong frustration of the
chains and their decoupling for $\Jp \ll 1$, a width of several
chains appears to give a good description of the physics of the
underlying two-dimensional lattice for smaller \Jp. We include a
careful reexamination of the zigzag chain, \ie width-2 cylinder,
which is then extended to wider systems. We do find an alternation in
the properties depending on whether the width is of the form $4n$ or
$4n+2$, with $n$ an integer, but this effect vanishes quickly with
increasing $n$. In particular for smaller couplings \Jp, we find that
our cylinders behave rather similarly to the zigzag chain. Overall,
we see incommensurate behavior over a wide parameter range for all
systems analyzed, with no indication of a collinear phase for smaller
\Jp.

This paper is thus organized as follows. Section I defines the model,
and reviews its classical phase diagram. Section II describes the
methods used to obtain incommensurate data, paying particular
attention to boundary conditions. Section III presents the results,
starting with a reexamination of the zigzag chain. This puts the
stage for the analysis of increasingly wider systems, followed by
summary and conclusions.

\subsection{The anisotropic triangular Heisenberg lattice }

The anisotropic triangular Heisenberg lattice is described by the
Hamiltonian
\begin{align}
  \hat{H}=\sum_{\left\langle i,j\right\rangle }
  J_{ij}\mathbf{\hat{S}}_{i} \cdot\mathbf{\hat{S}}_{j}
\text{,}\label{eq:H-heisenberg}
\end{align}
with the sum over all nearest neighbor pairs on the triangular
lattice, with $J_{ij}>0$
corresponding to frustrated antiferromagnetic (AF) nearest neighbor
interactions.
Dzyaloshinskii-Moriya interactions, which we do not include, are
expected to help stabilize the incommensurate phase analyzed in this
paper. \cite{Bode07,Ghamari11,Griset11} The strength of these
interactions may be, for example, on the order of a few percent of
$J$ for \CuCl. \cite{Coldea02}
The Hamiltonian in \Eq{eq:H-heisenberg} is depicted schematically in
\Fig{fig:lattice} in terms of a width-4 system. Here an $L \times \Ly$
system refers to $\Ly$ chains of length $L$ each. All energies are
expressed in units of $J$, leading to the single dimensionless
parameter $\Jp \equiv J^{\prime}/J$, with explicit reference to $J$
for emphasis only unless specified otherwise.

For practical reasons, the Hamiltonian in \Eq{eq:H-heisenberg} is
augmented by the additional term,
\begin{align}
  \hat{H}_{\mathrm{pin}} =
  \sum_{i} B_i^{\mathrm{pin}} \hat{S}_{i,z}
\text{,}\label{eq:H-pinning}
\end{align}
which describes pinning of a few sites $i$ at an open boundary.
These pinning fields (i) facilitate the numerical convergence, and
(ii) provide a particularly convenient way, for example, to calculate
and display complex correlations in a DMRG calculation. Regardless of
whether one sees incommensurate correlations through correlation
functions or through pinning, it is crucial that the boundary
conditions alter these correlations as little as possible. In
contrast, using periodic boundary conditions also along the
incommensurate chain direction would be particularly troublesome,
forcing commensurate locking and inducing sudden jumps in the
incommensurate wave vector. Therefore we avoid fully periodic
boundary conditions completely.

\subsection{Classical phase diagram}

The classical phase diagram of the anisotropic Heisenberg lattice at
zero temperature shows incommensurate order for the wide parameter
range $\Jp \in [0,2]$ due to the system's inherent frustration.
Within this parameter range, the classical ground state is given by a
spiral wave with the incommensurate wave vector $\qJ$ pointing along
the $J$-direction, $\vec{q} = \qJclass \hat{e}_J$.
\cite{Yoshimori59,Merino99} The classical spiral wave is defined as a
set of spins rotated in some arbitrary but fixed two-dimensional
plane by an angle $\vec{q}\cdot\vec{r}_{i}$ with $\vec{r}_{i}$ the
position of spin $i$ within the triangular lattice. Then for
arbitrary amplitude $q\equiv\qJclass$, the energy per site of the
spiral wave in the $J$-direction is given by
\[
   \Eclass (\Jp) =
   \cos\left(  q\right)  +2\Jp\cos\left(  \frac{q}
   {2}\right)  \text{,}
\]
having assumed spins of unit length, \ie $\left\vert S\right\vert
\equiv 1$, and lattice spacing $a \equiv 1$. This energy is minimized
by $ \cos( \tfrac{\qJclass}{2}) =-\frac{\Jp}{2}$ for $|\Jp|\le2$,
resulting in the classical ground state energy per site $\E0class$
for the incommensurate spiral wave with vector vector \qJclass given
by,
\begin{eqnarray}
   \qJclass ( \Jp ) &=& 2\bigl[  \pi -\cos^{-1}(  -\tfrac{\Jp}{2}) \bigr]
   \label{q_classical} \\
   \E0class(\Jp) &=& -1 - \tfrac{1}{2}\left( \Jp \right)^{2}
\text{,}\label{E0_classical}
\end{eqnarray}
for $\Jp \in [0,2]$. Here, $2\pi$ was added in \qJclass, so it lies
within the first Brillouin zone, while assuming the branch
$\cos^{-1}\left( x\right) \in\left[ 0,\pi\right] $. The pitch angle
$\theta$ of the spiral wave, \cite{White96} defined as the angle
between two spins at neighboring chains as one moves half a lattice
spacing along the chains, is given by $\theta(\Jp) =
\cos^{-1}(-\tfrac{\Jp}{2}) = \pi - \qJclass/2 \in
[90^\circ,180^\circ]$, with $\qJclass \in [0,\pi]$ for $\Jp \in
[0,2]$.

\begin{figure}[tb]
\begin{center}
\includegraphics[width=0.95\linewidth]{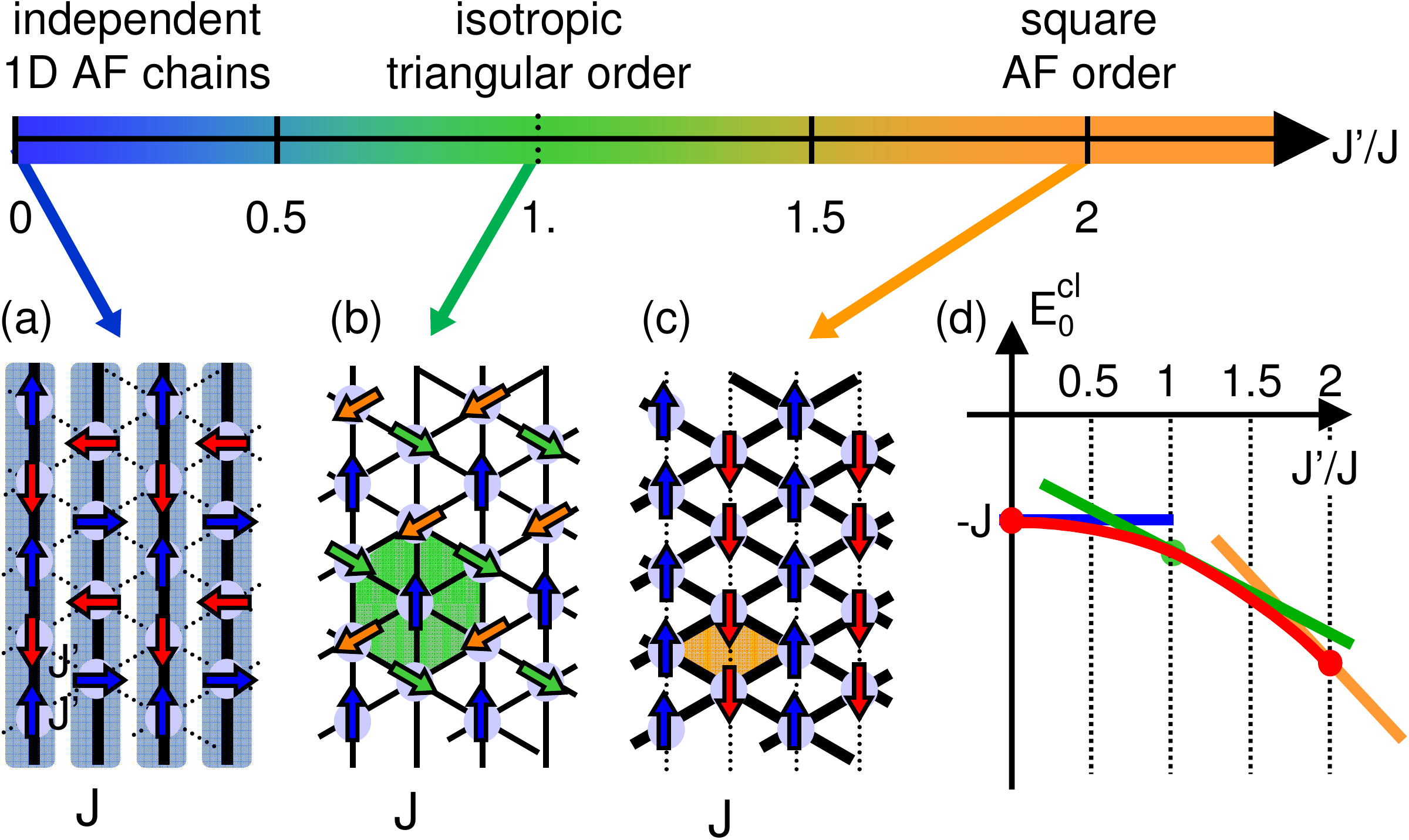}
\end{center}
\caption{
(Color online) Classical phase diagram of the anisotropic triangular
Heisenberg lattice at $T=0$ demonstrating continuous incommensurate
order for the entire interval $\Jp \in [0,2]$, having $J\equiv 1$.
Specific snapshots are shown for AF correlation for
$\Jp \to 0$ [panel a], triangular $120^{\circ}$ order at the
isotropic point $\Jp=1$ [panel b], and square AF
ordering for $\Jp \ge 2$ [panel c]. Panel d shows the classical
ground state energy per site for the spiral wave in $\Jp \in[0,2]$
[red (dark gray) line]. The three straight tangential lines around
the point $\Jp\in\{0,1,2\}$ assume the frozen spin configurations at these
points, respectively. For $\Jp \ge 2$, finally, the ground state
configuration is given by the commensurate square AF order. }
\label{fig_E0classical}
\end{figure}

The smooth classical incommensurate phase can be seen as the
continuous transition connecting the three commensurate points
$\Jp \in \{0,1,2\}$, as depicted in \Fig{fig_E0classical}.
(i) For small interchain coupling $\Jp \ll 1$, the chains are
essentially decoupled leading to antiferromagnetic spin correlation
along the chains (1D-AF), as indicated in \Figp{fig_E0classical}{a}.
Hence the incommensurate wave vector $\qJclass$ approaches the end of
the Brillouin zone of a single chain, \ie $\qJclass \to \pi$ [\cf
\Eq{q_classical}{}]. Note that with close to AF correlation within a
single chain, the interaction between chains is strongly frustrated
and hence suppressed. In particular, coinciding with our definition
of a spiral wave, the spins of a neighboring chain are displaced by
half a lattice constant and hence rotated by $\qJclass/2 = \pi/2$,
\ie $90^\circ$. The resulting \SS interaction across the chains is
thus close to zero, further emphasizing that neighboring chains
essentially decouple. Therefore in the frozen 1D-AF configuration,
$E/J=\left\langle S_{i} S_{i+1}\right\rangle =-1=\mathrm{const}$, as
indicated by straight line in \Figp{fig_E0classical}{d} around $\Jp =
0$.

(ii) At the isotropic point $\Jp = 1$, the system exhibits
$120^{\circ}$ order, as depicted in \Figp{fig_E0classical}{b}. The
wave vector of the spiral wave is given by $\qJclass =2\pi/3$, \ie a
period of three sites within a chain. If the order were frozen in the
$120^{\circ}$ structure, the energy per site would be
$E/J=-\frac{1}{2}\left( 1+2\Jp \right)$, as indicated by straight
line in \Figp{fig_E0classical}{d} around $\Jp = 1$.

(iii) For large interchain coupling $\Jp \gg 1$, the lattice reduces
to a square lattice along the $J^{\prime}$ couplings with weak
spin-coupling along one diagonal of the squares (diamonds) of
strength $J$, as indicated in \Figp{fig_E0classical}{c}. This leads
to a square AF order and consequently ferromagnetic (FM) order of the
spins along a single chain, \ie $\qJclass \to 0$. Within the frozen
square AF order, the ground state energy per site becomes
$E/J=1-2\Jp$, again indicated by a straight line in
\Figp{fig_E0classical}{d}. This square AF order is the true classical
ground state configuration for $\Jp \geq 2$ and agrees with
\Eq{E0_classical} for $\Jp=2$.

From a quantum mechanical point of view, this classical picture will
be altered by quantum fluctuations. Typically, one would assume that
quantum fluctuations will reduce incommensurate order. In particular,
while the phase boundary towards the square AF order also exists in
the quantum mechanical context, one expects that the incommensurate
phase terminates at a smaller value of \Jp, as compared to the
classical phase boundary of $\Jp=2$. For $\Jp<1$, however, the
question of whether or not quantum fluctuations fully suppress the
spiral wave into a collinear configuration for small enough yet
finite $\Jp$ has remained controversial. From our results below, we
do see clearly suppressed incommensurate order, in that the quantum
mechanical $\qJ$ approaches the boundary $\pi$ of the Brillouin zone
significantly faster as compared to the classical case. However, the
incommensurate correlations do persist for finite $\Jp$, suggesting
that $\qJ=\pi$ is reached only for $\Jp=0$.

\section{Methods}

We use the density matrix renormalization group (DMRG) \cite{White92}
on a finite two-dimensional lattice with mainly cylindrical boundary
conditions. We use \emph{traditional} DMRG in that a two-dimensional
strip of certain width is mapped onto a single effectively
one-dimensional chain, as indicated in \Fig{fig:lattice}. The
resulting ground state is therefore described by a matrix-product
state (MPS). \cite{Rommer95,Schollwoeck11} This approach provides a
numerically well-controlled setting, which, however, becomes
numerically expensive for smaller $\Jp$, and therefore prohibits a
fully converged analysis for $\Jp\lesssim 0.5$ for widths $\Ly>2$.
Nevertheless we are able to make a well-controlled and largely
unbiased analysis of the incommensurate correlations down to
$\Jp\gtrsim 0.5$.

\subsection{Cylindrical boundary conditions to study incommensurate correlations}

Incommensurate behavior is affected by boundary conditions imposed on
the finite system size under consideration, \cite{Misguich99} which
hence must be dealt with carefully. For this, we performed extensive
initial test calculations on the anisotropic triangular lattice with
a large variety of boundary conditions. For example, to allow any
type of incommensurate correlations to appear and not be frustrated,
we studied systems with fully open boundary conditions up to $11
\times 13$, with weak pinning of a single site in the center of the
system. All such calculations strongly indicated incommensurate
spiral correlations in the direction along the chains, varying with
$\Jp$. They also always gave a commensurate period of two chain
spacings ($\sqrt{3}a$) for transverse correlations, \ie
\emph{ferromagnetic} correlations in next-nearest neighbor chains.
\cite{Ghamari11}

Thus in order to study the incommensurate correlations in a
\emph{least} constrained way, we use cylindrical boundary conditions
(cyl-BC) with an even circumference, \ie composed of an even number
of chains [note that this is also compatible with the square AF order
of the system for large \Jp{}]. Furthermore, the very left boundary
of the open chains was pinned by a small external (staggered)
magnetic field, while the right boundary was \emph{softened} by
damping the Heisenberg couplings smoothly towards zero (\emph{smooth
boundary condition}). \cite{Vekic93} The resulting combined set of
boundary conditions will be referred to as cylindrical pinned with
smoothing boundary condition (\cpsBC). The pinning fields at the left
boundary induce an (exponentially) decaying magnetization in the bulk
of the system. The resulting incommensurate correlations are analyzed
away from the open boundaries in the central area of the system.

\subsubsection*{Finite size artifacts for small systems}

\begin{figure}[tb]
\begin{center}
\includegraphics[width=0.70\linewidth]{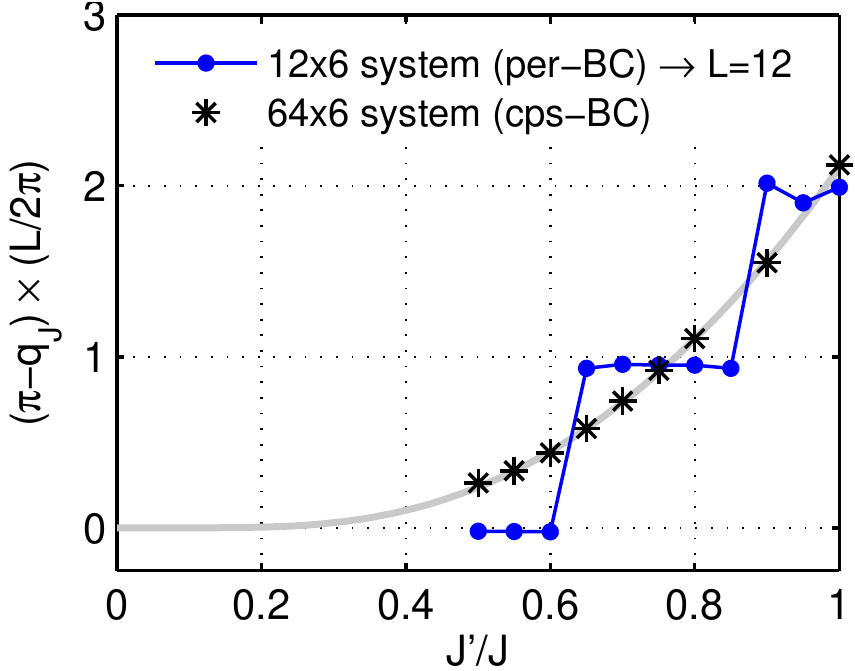}
\end{center}
\caption{
(Color online) Comparison of the incommensurate wave vector $\qJrel
\equiv \pi - \qJ$ obtained from DMRG between a small $12 \times 6$
system with fully periodic BC (\perBC) [solid line with round symbols]
and a larger $64 \times 6$ system using \cpsBC (data [black
asterisks] with a fit of the type $a (\Jp)^2 \exp{-b/\Jp}$
[gray line] taken from \Fig{fig:all_128x3} below). The incommensurate
data for the fully periodic system was extracted from the residual
$\Szx \sim 10^{-3}$ data derived from the calculated DMRG ground state
for $S_z^{\mathrm{tot}}=0$, consistent with explicit $\langle
S_0\cdot S_i\rangle$ correlation data. In the fully periodic system,
no pinning or smoothing was applied to guarantee full translational
invariance. Due to the presence of long-range
interactions, in the \perBC case up to $m=5000$ states had to be kept.
}\label{fig:qperBC} %
\end{figure}

Incommensurate correlations for $\Jp\lesssim 0.5$ exhibit
(exponentially) long wave lengths $\lambda\equiv 2\pi/(\pi-\qJ)$.
These correlations are strongly affected by small system sizes and
the boundary conditions applied, and as such may potentially be
misinterpreted. An example is given in \Fig{fig:qperBC}. For fully
periodic boundary conditions (\perBC), the relatively small $12
\times 6 $ system clearly shows finite size effects of the type
$\qJrel \equiv \pi - \qJ \simeq \tfrac{2\pi}{L} n$ with
$n=0,1,2,\ldots$ an integer. The small and noisy deviations from pure
integer $n$ may already be considered an indicator that the system
tries to break away from the periodicity enforced by given system
length $L=12$. In contrast, the incommensurate data for the larger
$64 \times 6 $ system, using \cpsBC clearly interpolates the \perBC
data in a smooth fashion. A fit of the form $\qJrel(\Jp) = a (\Jp)^2
\exp{-b/\Jp}$ is shown in \Fig{fig:qperBC} in solid gray [see also
\Fig{fig:all_128x3} later]. For the fully periodic system, even for
relatively large systems the transition between uniform collinear
behavior ($n=0$) and the first ``transition'' to $n=1$ will always
occur at relatively large $\Jp \gtrsim 0.5 $, which may thus be
misinterpreted as a transition into a collinear phase. Note that this
``transition'' changes the parity or reflection symmetry of the
ground state which has been used as an argument in favor of a
(possibly continuous) phase transition in the literature.
\cite{Heidarian09,Weng06}

In contrast, for all of our data using \cpsBC for as small as $\Jp
\simeq 0.3 \ldots 0.5$ for the width-4 system (not presented), we
still see incommensurate behavior, in that the magnetization data
shows a clear onset of oscillatory behavior consistent with our fit
to \qJ. It has significantly larger error bars, however, since (i)
many more states would actually have to be kept for full convergence
given that the entanglement block entropy strongly grows for smaller
\Jp, and (ii) the corresponding wavelength $\lambda = 2\pi/(\pi -
\qJ)$ can no longer be determined reliably as it clearly exceeds
accessible system sizes.

\subsection{Determination of the incommensurate wave vector}

\begin{figure}[tb!] \begin{center}
\includegraphics[width=1\linewidth]{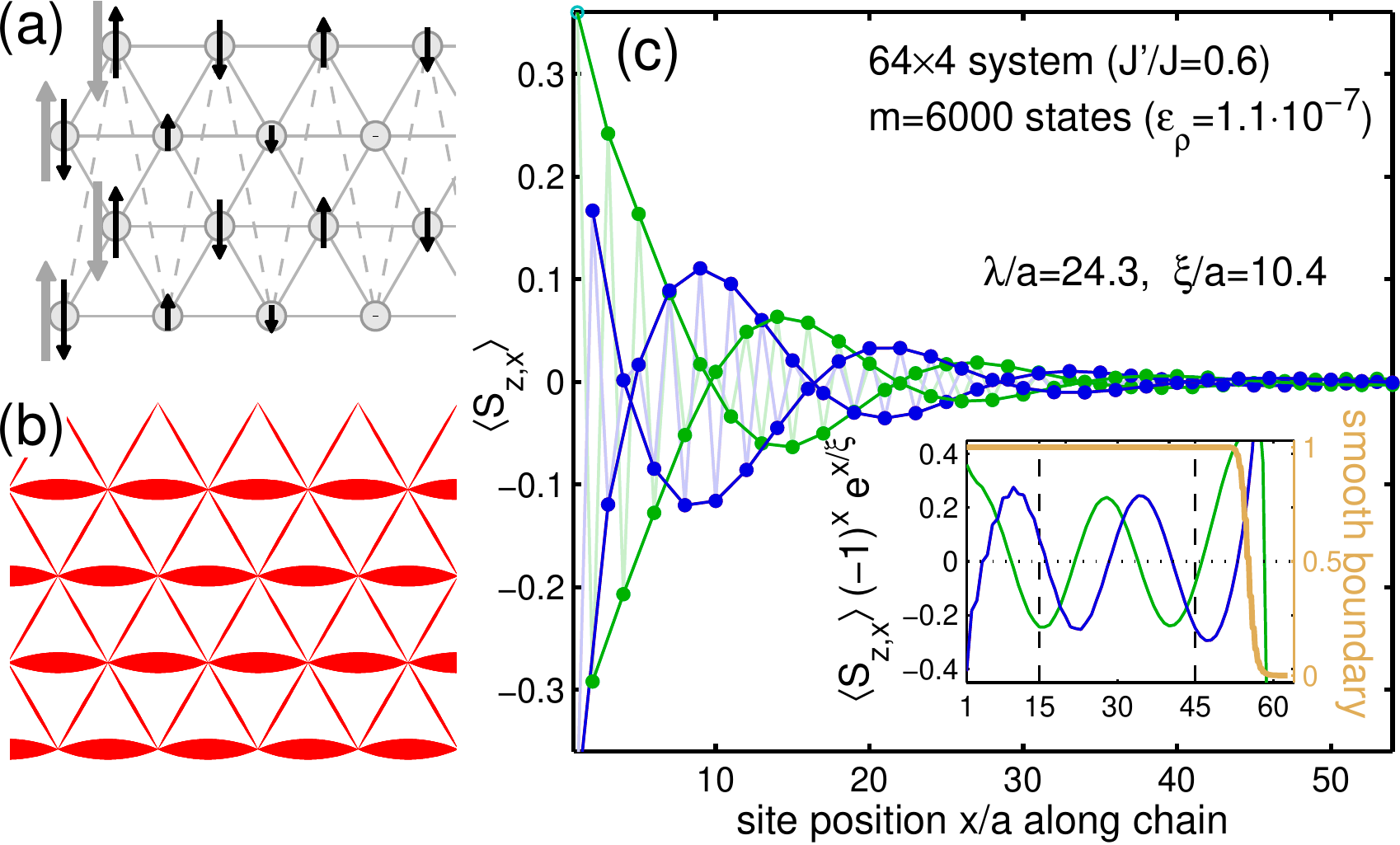}
\end{center}\caption{
(Color online) Analysis of the incommensurate correlations for
$64\times 4$ system at $\Jp=0.6$ using \cpsBC.
Panel (a) shows the magnetization $\Szx$ at the open left boundary
[black arrows on top of each site], as triggered by the staggered
pinning fields $B_{x,y}^{\mathrm{pin}}$ at the leftmost sites [gray
arrows] with $|B|=0.5$. The triangular lattice with sites and bonds
is indicated in the background, with the bonds due the periodic BC in
the vertical direction indicated by dashed lines.
Panel (b) shows $\langle S\cdot S \rangle$ correlations between
nearest-neighbor sites around the center
of the system. These correlations are well-converged, uniform,
and antiferromagnetic [indicated by the same red color], with
intrachain correlations $\langle S\cdot S \rangle_J \simeq -0.394$
and significantly weaker interchain correlations $\langle S\cdot S
\rangle_{J^\prime} \simeq -0.061$.
Panel (c) analyzes the full $\Szx$ response of the system, as partly
already indicated in panel (a), as a function of horizontal
position for all chains. It shows the bare $\Szx$ data [light
colors], together with the exponentially decaying oscillating
envelopes [strong colors], from which the exponential decay $\xi$
and the incommensurate period $\lambda$ are determined from a phase
analysis, as described in \Eq{eq:CosSin} and the following
discussion. The inset shows the reduced purely oscillating part
of $\Szx$. The right axis set of the inset and
its corresponding data [matching colors] indicate the weights applied
to the couplings for smoothing the open right boundary.
} \label{fig:SZanal} %
\end{figure}

\begin{figure}[tb!] \begin{center}
\includegraphics[width=1\linewidth]{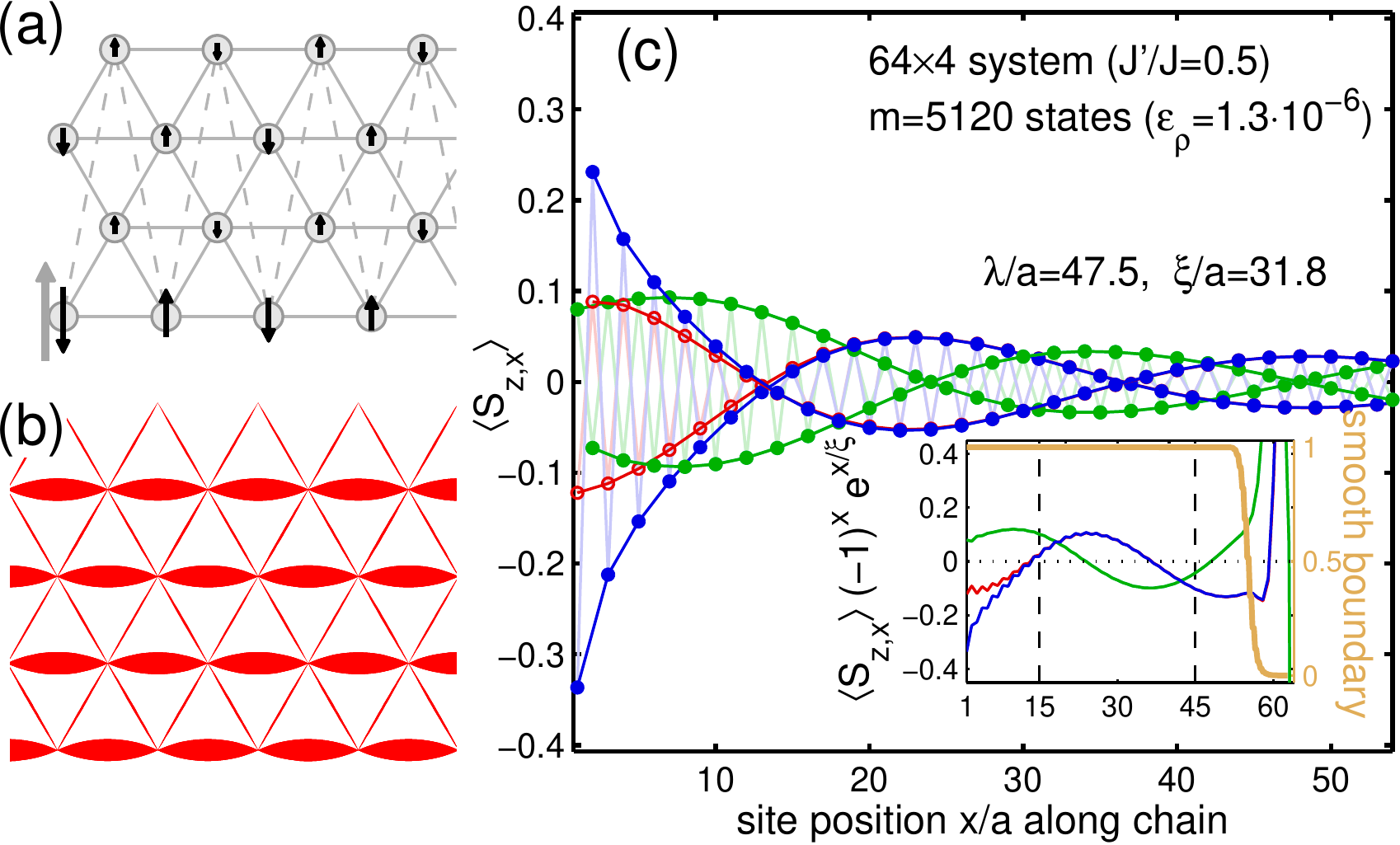}
\end{center} \caption{
(Color online) Analysis of the incommensurate correlation for the
same system as in \Fig{fig:SZanal}, except for smaller $\Jp=0.5$ and
the pinning field which is applied to a \emph{single} site a the left
boundary only [indicated by the light gray arrow in panel (a);
B=0.5]. Panel (b) shows the intrachain correlations at the center of
the system, having $\SSJ \simeq -0.419$ with strongly weakened
interchain correlations $\SSJp \simeq -0.038$ due to frustration.
Note that \SSJ is already close to the lower bound for the mean of
$\SSJ \ge \tfrac{1}{4}-\ln(2) = -0.4431$, derived from the ground
state energy $E_0/J$ of a single Heisenberg chain. \cite{Oliveira93} }
\label{fig:SZanal:LH1}
\end{figure}

The incommensurate wave vector is determined by the analysis of the
system's response to the pinning fields at the left boundary using
\cpsBC. The procedure is illustrated for a $64\times 4$ system for
$\Jp=0.6$ in \Fig{fig:SZanal}, and with altered pinning for $\Jp=0.5$
in \Fig{fig:SZanal:LH1}. Note that despite $\Jp\simeq 0.6$ was
suggested as the phase boundary towards collinear order,
\cite{Heidarian09} both systems, \Fig{fig:SZanal} as well as
\Fig{fig:SZanal:LH1}, clearly show pronounced incommensurate
oscillations still, while having $\Jp \le 0.6$.

Using \cpsBC, in \Figp{fig:SZanal}{a} the leftmost site of each chain
is pinned through a staggered external magnetic fields
$|B_{\mathrm{pin}}|=0.5$ which thus respects the underlying AF
correlations of the Heisenberg model for smaller \Jp. However, the
exact details of the applied pinning usually did not matter [see
\Fig{fig:SZanal:LH1} later]. After a relatively short transient
region, the magnetization of each chain followed a clear exponential
decay with superimposed oscillations, as seen in
\Figp{fig:SZanal}{c}. The period of these oscillation usually neither
is a simple multiple of the underlying lattice spacing $a$, nor does
a multiple of the period fit into the specific finite system size
under investigation, \ie the period is incommensurate. The
\emph{smoothing} \cite{Vekic93} of the right open boundary roughly
affected the right 20\% of the system [see data associated with right
axis in inset to \Figp{fig:SZanal}{c}]. Within the smoothing region
at the right boundary, both couplings, $J$ as well as $J'$, were
damped uniformly as a function of horizontal chain position $x$ by
weights that smoothly turned into an exponential decay $\propto
e^{-\Lambda x}$, \ie decreasing the couplings by a factor of
$\Lambda=2$ within one horizontal lattice spacing $a$. This setting
has been used for smooth boundary throughout. The purpose of this
smooth boundary in the \cpsBC setup was tailored to blur the finite
size in the direction of the chains, and hence to least constrain
incommensurate correlations.

The incommensurate correlations for smaller \Jp then are dominated by
AF correlations, as the wave vector $\qJ$ rapidly approaches the
boundary of the Brillouin zone of a single chain, $\qJ \to \pi$. This
is seen in the zigzag structure of the bare $\Szx$ data for $\Jp=0.6$
in \Figp{fig:SZanal}{c} [light colors in the background], while the
envelope for every other site [lines in strong color] are plain
decaying oscillating $\pm\sin()$ and $\pm\cos()$ curves for even and
odd chains, respectively. Note that the data for all even or odd
chains in \Figp{fig:SZanal}{c} coincides, and hence lies
indistinguishable on top of each other.

The spiral correlations are analyzed then as follows. With a
two-chain periodicity normal to the chains, the system can be
regarded as an interleaved set of even chains (chains $2,4,\ldots$)
and odd chains (chains $1,3,\ldots$). Consequently, the position $x$
of the sites in chain direction in the odd chains [$x = \tfrac{1}{2},
\tfrac{3}{2}, \ldots$, in units of lattice spacings $a$] is shifted
by half a lattice constant with respect to the even chains
[$x=1,2,\ldots$]. With $\Szx$ the measured spin projections in
z-direction of the spin at site position $x$, the exponentially
decaying envelope $\langle S_{z,0} \rangle \exp{-x/\xi}$ allows to
determine the correlation length $\xi$ by fitting. With $\qJrel
\equiv \pi -q_{J} \ll 1$ quickly becoming small for $\Jp<1$, the pure
oscillatory part of the spiral correlations along the chains can be
extracted. Up to an irrelevant overall phase, it is given by
\begin{align}
    \Szx &/ (\langle S_{z,0}\rangle \exp{-x/\xi}) \sim
       \cos \left( \left( \pi -\qJrel\right) x\right)  \label{eq:CosSin} \\
    =&\cos \left( \pi x\right) \cos \left( \qJrel x\right) +\sin \left(
    \pi x\right) \sin \left( \qJrel x\right)  \nonumber\\
    =&\left\{
       \begin{array}{ll}
       (-1)^{x} \cos \left( \qJrel x\right)
         & \text{for } x=1,2,\ldots \text{ (even chains)} \nonumber\\
       (-1)^{\tilde{x}} \sin \left( \qJrel x\right)
         & \text{for } x=\tfrac{1}{2},\tfrac{3}{2},\ldots  \text{ (odd chains)}\text{,}
       \end{array}
    \right. 
\end{align}
with $\tilde{x} \equiv x-\tfrac{1}{2}$ in the last line. This zigzag
due to the signs together with the oscillatory envelope of sine and
cosine waves is clearly seen in the main panel \Figp{fig:SZanal}{c}.
Here the global phase is fixed through the pinning at the left
boundary, thus facilitating the overall numerical convergence within
the DMRG calculation. By applying staggered signs and correcting for
the overall exponential decay, pure cosine (even chains) and sine
waves (odd chains) can be extracted, as shown in the inset to
\Figp{fig:SZanal}{c}. Here the sign-factor for odd chains needs to be
understood as $(-1)^{\tilde{x}}$, as introduced with \Eq{eq:CosSin}.
The incommensurate wavelength $\lambda \equiv 2\pi/(\pi-\qJ)$ of the
slowly oscillating envelope can then be determined, for example, from
the zero-transitions of these oscillations, assuming that several
periods fit into the system.

Alternatively, a phase analysis of the the cosine-sine relationship
in \Eq{eq:CosSin} can be employed to determine $\qJ$. For this, note
that away from the open boundaries, the slow oscillations of the
envelope in \Figp{fig:SZanal}{c} or its inset are well described by
$c(x)\equiv r(x) \cos(\varphi(x))$ and $s(x)\equiv r(x)
\sin(\varphi(x))$, with $\varphi(x)\equiv \qJ x$, up to an irrelevant
overall phase, and a common decaying envelope function $r(x)$. Here
even and odd chains are only distinguished by their respective
discrete sets of values for $x$. Nevertheless, for example, by
interpolating the sine data for odd chains half-way in between two
neighboring sites, values $(c(x),s(x))$ for a matching position $x$
are obtained. With $\tan(\varphi(x)) = s(x)/c(x)$, the wave vector
\qJ can thus be determined from the slope of the calculated phase
$\varphi(x)$. The amplitude $r(x)$ drops out, hence its precise value
and functional dependence is unimportant. This phase analysis,
indeed, represented a reliable alternative procedure to determine \qJ
for smaller \Jp. In particular, it also showed the quality of the
underlying sine and cosine data, which for the systems in
\Fig{fig:SZanal} or \Fig{fig:SZanal:LH1} demonstrated an excellent
linear dependence of $\varphi(x)$ over the fitting range $x$
indicated by the vertical dashed lines in the inset to panels (c).
The specific resulting values for the exponential decay $\xi$ and the
wavevector \qJ are specified with the panel.

The analysis in \Fig{fig:SZanal} has been repeated for exactly the
same system, yet for smaller $\Jp=0.5$ and with the pinning reduced
to a \emph{single} site (ssp) at the left boundary, as indicated in
\Figp{fig:SZanal:LH1}{a}. If the same $\Jp=0.6$ as in
\Fig{fig:SZanal} had been taken, the altered pinning of
\Fig{fig:SZanal:LH1} solely resulted in a modified transient behavior
right next to the pinning fields at the left boundary, which also
leads to a different irrelevant phase of the oscillatory part in
$\Szx$. The resulting correlation length $\xi$ as well as the
incommensurate wave vector $\lambda$, however, are exactly the same
as already indicated in \Figp{fig:SZanal}{c}, with relative
differences on the order of 1\%. This insensitivity of the
incommensurate behavior to the exact details of the pinning at the
left boundary is seen also for a wider range of $\Jp$, as will be
demonstrated in \Fig{fig:all_128x2}.

The analysis in \Fig{fig:SZanal:LH1} then is based on a system with
the smaller interchain coupling $\Jp=0.5$, instead. The pinning
occurs on a single site at the lowest chain, considered chain \#1,
and hence an odd chain. Similar to \Fig{fig:SZanal}, in the main
panel \Figp{fig:SZanal:LH1}{c} a transient behavior at the left
boundary is clearly visible. Not surprisingly, the data within the
odd chains differs for $x/a\lesssim 15$, given that one of them is
pinned. Overall, however, data for even or odd chains quickly
coincide away from the left boundary, consistent with what has
already been seen in \Fig{fig:SZanal}. Also, the data for even chains
coincides from the very beginning. This is attributed to the very
weak \SSJp correlation in between the chains [see
\Figp{fig:SZanal:LH1}{b}] due to the systems inherent frustration
despite the sizeable \Jp of 0.5.

\section{Results}

\subsection{Review of width-2 system (zigzag chain) \label{sec:width2}}

The triangular system consisting of two chains is also referred to as
zigzag or $J_1$-$J_2$ chain, with nearest-neighbor interaction $J_1
\equiv J'$ and next-nearest neighbor interaction $J_2 \equiv J$.
While it has been widely studied in the literature,
\cite{Vekic93,Eggert96} we carefully reexamine the zigzag chain in
the entire parameter range from small to large $\Jp$, with the main
focus on incommensurate behavior \cite{White96} for $\Jp<1$. This
analysis for the width-2 system then sets the stage for the wider
systems further below, which will proceed in a completely analogous
fashion.

\begin{figure}[tbp!]
\begin{center}
\includegraphics[width=1\linewidth]{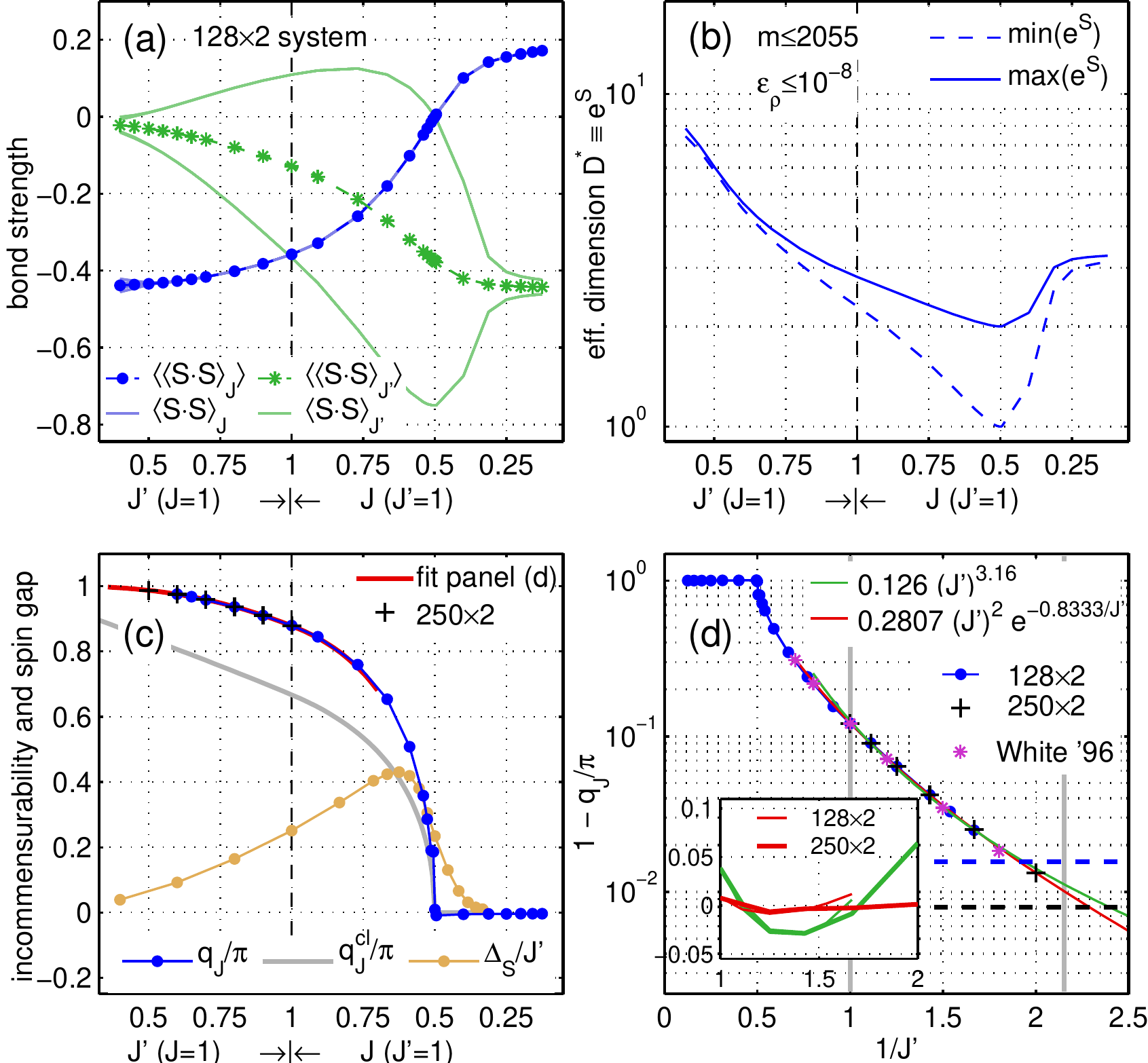}
\end{center}
\caption{
(Color online) Analysis of $128\times 2$ system (zigzag chain)
around the system center using \cpsBC{} over a wide range of $\Jp$.
In panels (a-c), the horizontal axis shows \Jp for $\Jp<1$,
smoothly switching to the inverse $1/\Jp$ for $\Jp>1$ [\cf
\Eq{eq:xScale}{}].
Panel (a) shows the nearest-neighbor spin correlations of
individual bonds along the chains [$\SSJ$] and in between the
chains [$\SSJp$]. Here, up to dimerization, this leads to many
lines lying on top of each other, with minor deviations
seen for the smallest \Jp only.
\SSJa and \SSJpa corresponds to the averaged $\SSJ$ and
$\SSJp$ data, and is shown in strong colors [dashed with bullets
and asterisks, respectively].
Panel (b) indicates the numerical cost of the calculations in terms
of the effective dimension $\Deff\equiv\exp{S}$ (see text). \Deff is
calculated \wrt to bonds of the linearized system [\cf path shown in
\Fig{fig:lattice}]. Given intrinsic even-odd alternations, for
simplicity, only the maximum and minimum \Deff from block-decompositions
\wrt the system center is shown.
Panel (c) shows the incommensurate wave vector $\qJ$ [solid blue
with bullets for $128\times2$ system, black pluses for larger
$250\times2$ system], where the thick solid line for smaller \Jp
replicates the exponential fit from panel (d). For reference, also
the classical incommensurate wave vector \qJclass as well as the
spin-gap $\Delta_S/J$ is shown, with the latter calculated for
plain cylindrical BC (see text).
Panel (d) analyzes the incommensurate data $\qJ$ for small \Jp
relative to the zone boundary vs. plain inverse \Jp on a
semilogarithmic plot. A smooth exponential fit [solid red (dark
gray)], and for comparison, a plain polynomial fit [solid green
(light gray)] are shown. Data for a larger $250\times 2$ system as
well as data from White~'96 [\onlinecite{White96}] are included. The
horizontal dashed lines indicate $2/L$, \ie the smallest $\qJrel/\pi
\equiv1-\qJ/\pi$ reachable for given system size [color match with
data in panel]. The inset shows the relative deviation of both fits
from the data. Thin (thick) lines are for the smaller (larger)
system, while red (dark gray) [green (light gray)] lines refer to
the exponential [simple polynomial] fit, respectively.
} \label{fig:all_128x1}
\end{figure}

The results for the $128 \times 2$ system are summarized in
\Fig{fig:all_128x1} using \cpsBC. Since for the zigzag chain the
periodic boundary in the width of the system is equivalent to taking
$\Jp \to 2\Jp$ and using open BC, the boundaries are considered open
in this case, while nevertheless applying pinning and smoothing as
usual. The data shown in \Fig{fig:all_128x1} covers a wide range of
$\Jp$ from large $\Jp \gg 1$ down to smaller $\Jp \gtrsim 0.5$. For
this purpose, panels (a-c) plot the data vs. $J^{\prime}$ in units of
$J$ for $\Jp \le 1$, while for $\Jp>1$ the data is plotted vs. $J$ in
units of $J^{\prime}$ in reverse order. To be specific, while $J$ and
$J'$ is indicated on the horizontal axis in panels
\Figp{fig:all_128x1}{a-c} for readability, what is actually plotted
on the horizontal axis is
\begin{align}
   \x \equiv \left\{
   \begin{array}{ll}
      \tfrac{J^{\prime }}{J} & \text{for }J^{\prime }/J\leq 1 \to \x\in[0,1] \\
      2-\tfrac{J}{J^{\prime }} & \text{for }J^{\prime }/J\geq 1 \to \x\in[1,2]
   \end{array}
\right.\label{eq:xScale}
\end{align}
Overall then, $\x \in [0,2]$ covers the entire range $\Jp \in
[0,\infty]$, with $\x=1$ being the isotropic triangular lattice. Note
that the derivative of $\x(\Jp)$ is smooth across $\Jp=1$, which is
also reflected in the smoothness of all data across $\Jp=1$ in panels
\Figp{fig:all_128x1}{a-c}.

Panel (a) of \Fig{fig:all_128x1} analyzes the nearest-neighbor
correlations \SSJ and \SSJp at the center of the system along and in
between the chains, respectively. The overall averages \SSJa and
\SSJpa are shown in strong solid colors with symbols. The data for
individual bonds, \SSJ and \SSJp, with respect to $\nc=8$ sites from
both chains is shown in light colors [solid lines]. Much of the data
of individual bonds lies indistinguishable on top of each other,
which demonstrates the uniformity of the system [larger deviations
will be seen later for wider systems for small \Jp due to numerical
issues [\eg see \Fig{fig:all_128x3}{}]. In \Fig{fig:all_128x1}, tiny
deviations in the individual bond data are seen only for the very
smallest $\Jp=0.4$ analyzed. Despite numerical issues as discussed
with panel (b) below, this is also attributed to finite size effects,
in that the incommensurate wavelength $\lambda \equiv 2\pi/(\pi-\qJ)$
reaches and rapidly extends beyond given system size for small $\Jp$.

While the \SSJ data widely agrees with its average, the \SSJp data
shows a symmetry-broken state. The interchain bonds combine two
different diagonal directions, and as such shows dimerization over a
wide range, \cite{White96} seen as the opening of a
\emph{dimerization bubble} in the \SSJ data. This bubble closes, \ie
approaches its average [asterisks] for $\Jp\to 0$ and for large \Jp
at $1/\Jp \simeq 0.241167$. \cite{Eggert96} The dimerization results
from spontaneous symmetry breaking along the direction of the chains
with alternating weak and strong interchain bonds [interestingly, a
similar symmetry breaking is encountered again later in an
increasingly weaker form for the wider width-6 and width-10 systems].
The width-2 system analyzed here becomes completely dimerized at the
Majumdar-Ghosh point, \cite{Majumdar69} $\Jp=2$, as seen in
\Figp{fig:all_128x1}{a} at $J/\Jp=0.5$. There both, the \SSJ data
[blue (dark gray) line with bullets] as well as the upper branch in
the \SSJp data [solid light green (gray) lines], pass through zero,
while the lower branch in the \SSJp data reaches its strongest
negative value of $-0.75$ due to pairwise singlet formation.

The numerical cost of a DMRG calculation is directly reflected in the
effective dimension $\Deff \equiv \exp{S}$, which is plotted in
\Figp{fig:all_128x1}{b}. Here $S$ is the block-entropy around the
center of the system, \ie the von-Neumann entropy after tracing out
approximately half of the system. Up to a prefactor, the effective
dimension \Deff directly indicates the dimension $D$ of the
underlying matrix product state that is required for some fixed
prespecified accuracy. As such, \Deff indicates the numerical cost,
which in the case of DMRG scales as $\mathcal{O}(D^3)$.
For reference, \Figp{fig:all_128x1}{b} also indicates the actual
number of states [$m \le 2055$, largest for small $\Jp$] as well as
the maximum discarded weight, $\epsilon_\rho$. \Deff typically shows
even-odd behavior and also variations depending on the explicit
block-partitioning of the system. Hence the maximum and minimum \Deff
across the system center is shown. As seen in
\Figp{fig:all_128x1}{b}, \Deff saturates for large \Jp, and exhibits
a minimum at the Majumdar-Ghosh point, $\Jp=2$. There \Deff
alternates between the minimum of 1 [at the boundary in between two
singlets] and the maximum of 2 [cutting across one singlet]. Starting
from the Majumdar-Ghosh point, when decreasing \Jp, \Deff increases
exponentially, with a further strong boost for $\Jp \lesssim 0.6$
[note that panel (b) is a semilogarithmic plot]. The strong increase
in numerical cost for small $\Jp$ is clearly due to the effective
decoupling of the chains in this parameter regime. This leads to
largely independent Hilbert spaces that need to be combined in a
tensor product. Nevertheless, the presence of the frustrating
neighboring chains does affect the detailed nature of the effective
low-energy Hilbert spaces, hence the sweeping path across the chains
as depicted in \Fig{fig:lattice} is important, and cannot simply be
replaced, for example, by a sweep preferentially along entire chains
first.

The results for the incommensurate wave vector $\qJ$ are shown in
\Figp{fig:all_128x1}{c} [blue (dark gray) bullets], together with
data from a larger $250\times 2$ system [black pluses] and an
exponential fit for small \Jp, replicated from panel (d) [thick red
(black) line]. The incommensurate wave vector $\qJ$ vanishes at the
Majumdar-Ghosh point, being zero for $\Jp \ge 2$. This phase boundary
incidentally agrees with the classical incommensurability \qJclass
for the infinite system. On the other hand, while for small $\Jp$ the
classical \qJclass approaches the boundary of the Brillouin zone in a
linear fashion [also plotted in panel (c) for comparison], the
quantum mechanical incommensurability is strongly reduced, in that
$\qJ$ approaches the zone boundary of $\pi$ much faster, and at first
sight, even appears to vanish already for $\Jp \simeq 0.5$. But as we
will argue in the following, it does not.

The spin-gap $\Delta_S$ of the zigzag chain [also calculated and
shown in panel c, for reference; see later discussion] is described
for small $\Jp$ by $\Delta_S \simeq c_1 \exp{-c_2/\Jp}$,
\cite{White96} with constants $c_1$ and $c_2$ of order one. For large
\Jp, on the other hand, the dimerization [panel (a)] as well as the
spin-gap [panel (c)] are expected to vanish for $1/\Jp = 0.241167$.
\cite{Eggert96} Motivated by this inverse exponential behavior of the
spin-gap for small \Jp, \Figp{fig:all_128x1}{d} shows the $\qJ$ data
of panel (c) vs. plain inverse \Jp. Moreover, in order to zoom into
the boundary of the Brillouin zone, the incommensurate data \qJ is
plotted in terms of $\qJrel \equiv \pi-\qJ$ on a semilogarithmic
scale in y-direction. Clearly, the incommensurate \qJrel decays fast
for large x-values [\ie small \Jp values], close to exponentially,
indeed, but by no means does \qJrel show any tendency to vanish for
finite $\Jp$. On the contrary, the data shows a slight upward
curvature.

We fitted the data for $\qJrel$ in the interval indicated by the two
vertical lines in \Figp{fig:all_128x1}{d} in two ways: (i) an
exponential fit of the type
\begin{align}
   \qJrel(\tfrac{1}{\Jp}) &\equiv \pi - \qJ(\tfrac{1}{\Jp}) \nonumber\\
&\cong c_1 (\Jp)^{c_3} \exp{-c_2/\Jp}
\text{,}\label{eq:qJexpfit}
\end{align}
and (ii), for comparison, also a plain polynomial fit. The
exponential fit indicated an exponent $c_3 \simeq 2$, so $c_3$ was
fixed to this value for the zigzag chain. The remaining fit
parameters are shown in the legend of panel (d). For comparison, the
plain power law fit results in $(\Jp)^{3.16}$, in agreement with the
$\mathcal{O}(\Jp{}^3)$ estimate by [\onlinecite{Ghamari11}]
in the case where spiral order is selected by fluctuations
at $\mathcal{O}(\Jp{}^2)$. It is
hard to discern in panel (d), which of the two fits is closer to the
data, so the \emph{relative} difference of the actual data to the
fitted values is shown in an inset to panel (d). The slight positive
curvature of the power-law fit in the panel appears somewhat too
strong, which is clearly magnified still in the inset. In comparison,
the exponential fit lies significantly closer to the actual data,
which due to the large number of states kept in the calculation, is
well-converged.

From this we conclude, that the exponential fit of the type $c_1
(\Jp)^2 \exp{-c_2/\Jp}$, which is non-analytic in $\Jp$, fits best
for the incommensurate wave vector of the zigzag chain. Moreover,
from the systematic behavior seen in the incommensurability down to
$\Jp\gtrsim 0.5$, we take this as a strong indication that $\pi-\qJ$
\emph{remains} finite for any finite $\Jp<0.5$. From further
calculations for $\Jp\sim 0.3\ldots 0.5$ (not shown) we do see that
the oscillatory bending of the $S_z$ data as in \Fig{fig:SZanal}
continues. The system, however, can no longer be taken large enough
to accommodate even a single full period of an incommensurable wave,
which would allow a reliable determination of $\qJ$. Clearly, given
the exponentially rapid decay of $\pi-\qJ$ as in $\exp{-c_2/\Jp}$,
the required system sizes to actually analyze incommensurable order
for small $\Jp$ becomes exponentially large. With the fit parameters
in panel (d), for example, the required system length estimated by
$\lambda \equiv 2\pi/\qJrel$ for $\Jp=0.3$ is around $\lambda \simeq
1,300$ sites, while for $\Jp=0.2$ it would have already grown to
$\lambda \simeq 11,500$ sites!

\subsection{Width-4 to width-10 systems}

\begin{figure}[tb!]
\begin{center}
\includegraphics[width=1\linewidth]{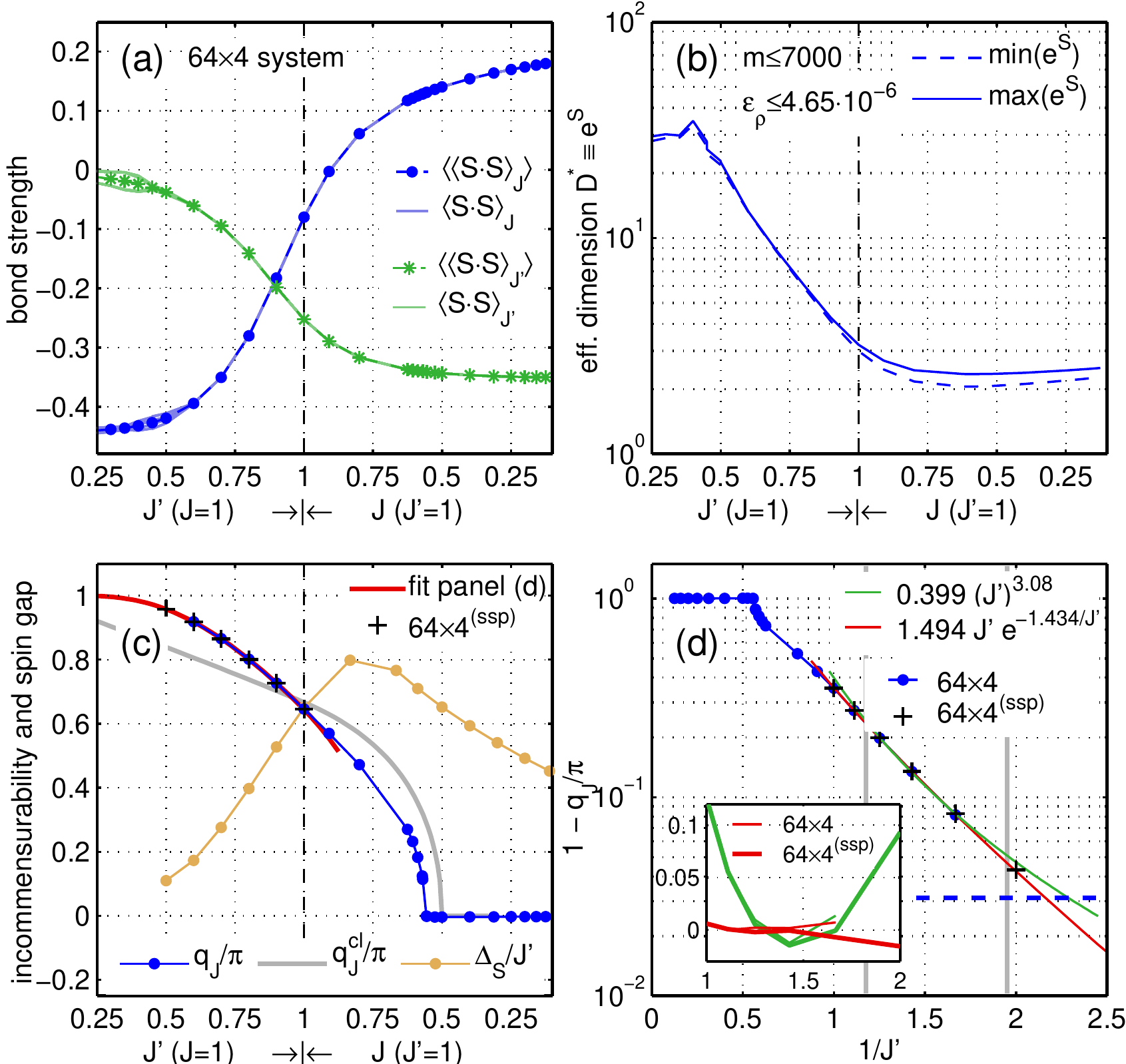}
\end{center}
\caption{(Color online)
Analysis of $64\times 4$ system using \cpsBC [analysis is similar to
\Fig{fig:all_128x1}; for a detailed description of panels and insets
see caption there]. The system shows no dimerization, with the
incommensurate phase boundary at $\Jpc\simeq 1.78$. Small
finite-size and numerical limitations are seen for $\Jp < 0.5$ in
panels (a-b). The exponential fit in panel (d) as in \Eq{eq:qJexpfit}
gives $c_3 \simeq 1$ to a good approximation, hence $c_3$ has been
fixed to $1$. The reference data [black crosses] shown in panels (c)
and (d) derive from exactly the same physical system, with the only
difference of having a single site pinned only [\cf
\Fig{fig:SZanal:LH1}]. } \label{fig:all_128x2}
\end{figure}

\begin{figure}[t!]
\begin{center}
\includegraphics[width=1\linewidth]{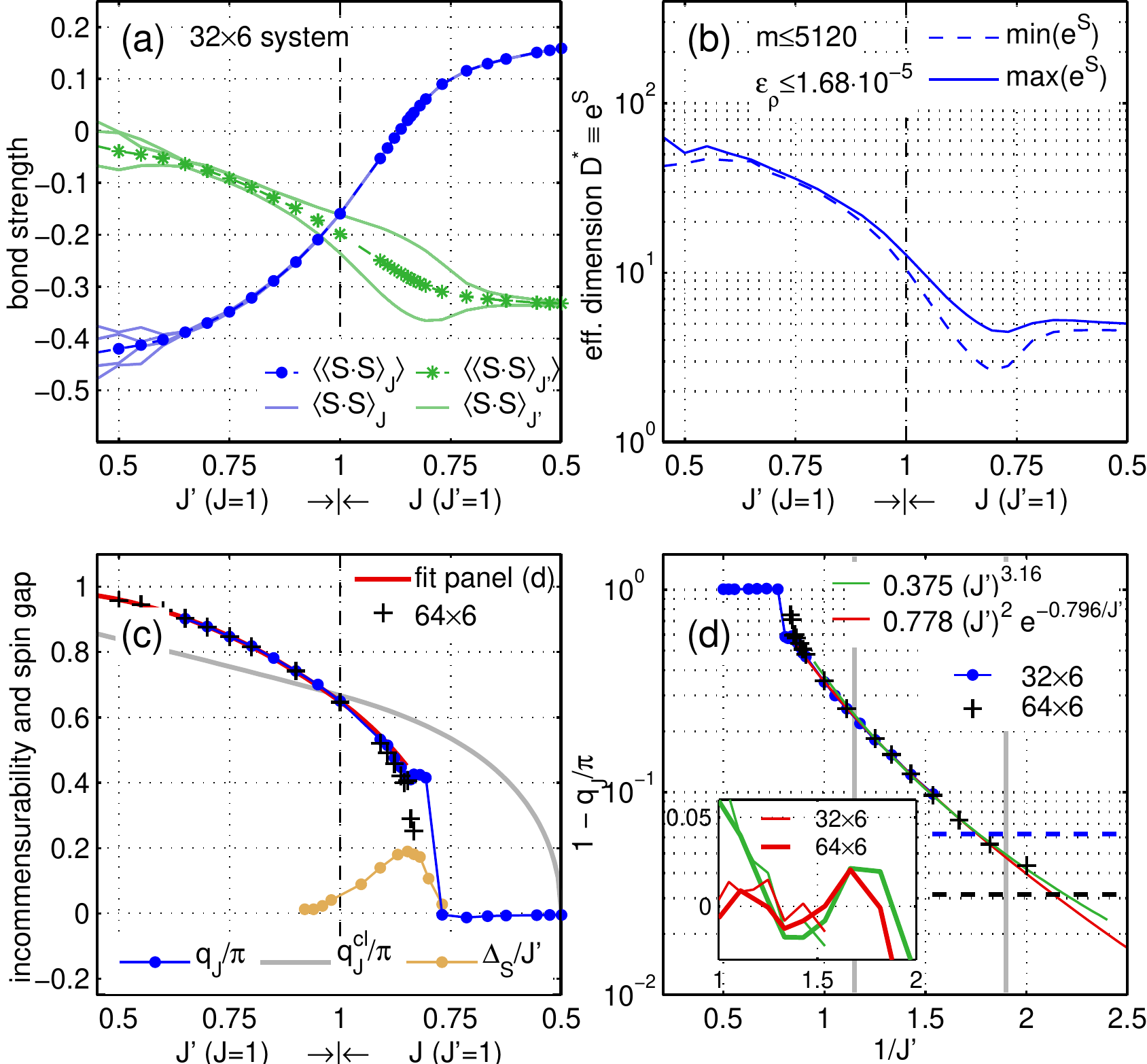}
\end{center}
\caption{(Color online)
Analysis of $64\times 6$ system using \cpsBC [analysis is similar to
\Fig{fig:all_128x1}; for a detailed description of panels and insets
see caption there]. The system again shows spontaneous symmetry
breaking, with the associated dimerization pattern at $\Jp=1.16$
shown in \Fig{fig:dmz}. Strong finite-size and convergence issues are
seen for $\Jp\lesssim 0.6$ in panels (a-b). The phase boundary for
incommensurate behavior (panel c) is given by $\Jp\le \Jpc \simeq
1.27$. Similar to the width-2 system, the exponential fit as in
\Eq{eq:qJexpfit}{} in panel (d) results in $c_3 \simeq 2$, thus
$c_3$ has been fixed to this value.
} \label{fig:all_128x3}%
\end{figure}

\begin{figure}[t!]
\begin{center}
\includegraphics[width=1\linewidth]{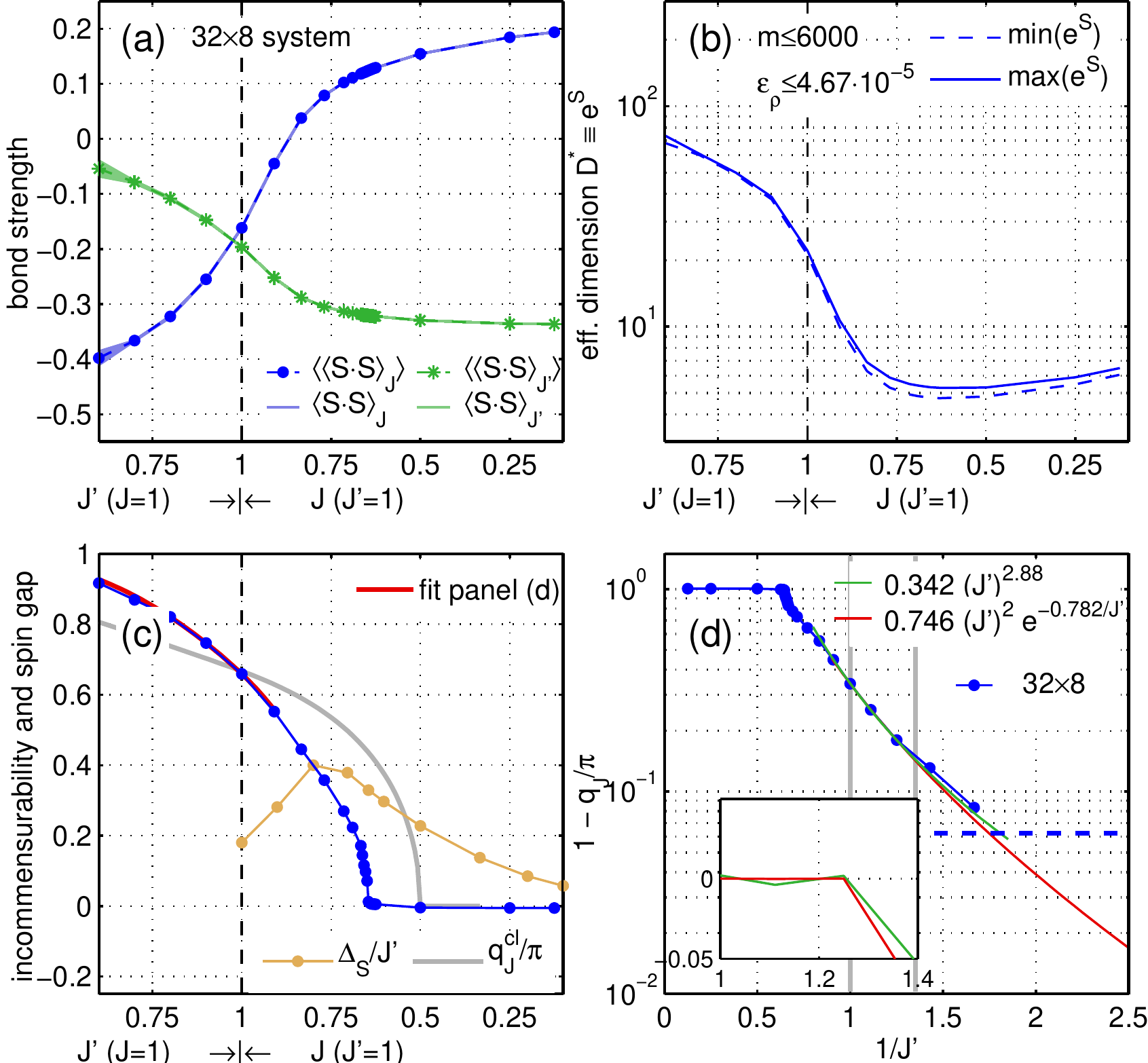}
\end{center}
\caption{(Color online)
Analysis of $64\times 8$ system using \cpsBC  [analysis is similar to
\Fig{fig:all_128x1}; for a detailed description of panels and insets
see caption there]. The system is uniform without any spontaneous
symmetry breaking, with incommensurate behavior for $\Jp\le \Jpc
\simeq 1.56$. Finite-size and convergence issues are seen for
$\Jp\lesssim 0.6$ in panel (a), with significant numerical truncation
starting with $\Jp\lesssim 0.8$, as indicated by the artificial
suppression (kink) of \Deff in panel (b). The exponential fit in
panel (d) uses $c_3=2$ [\cf \Eq{eq:qJexpfit}{}], although
the fitting range no longer supports a clear preference for either
$c_3=1$ or $c_3=2$.
}\label{fig:all_64x4}%
\end{figure}

\begin{figure}[tbp!]
\begin{center}
\includegraphics[width=1\linewidth]{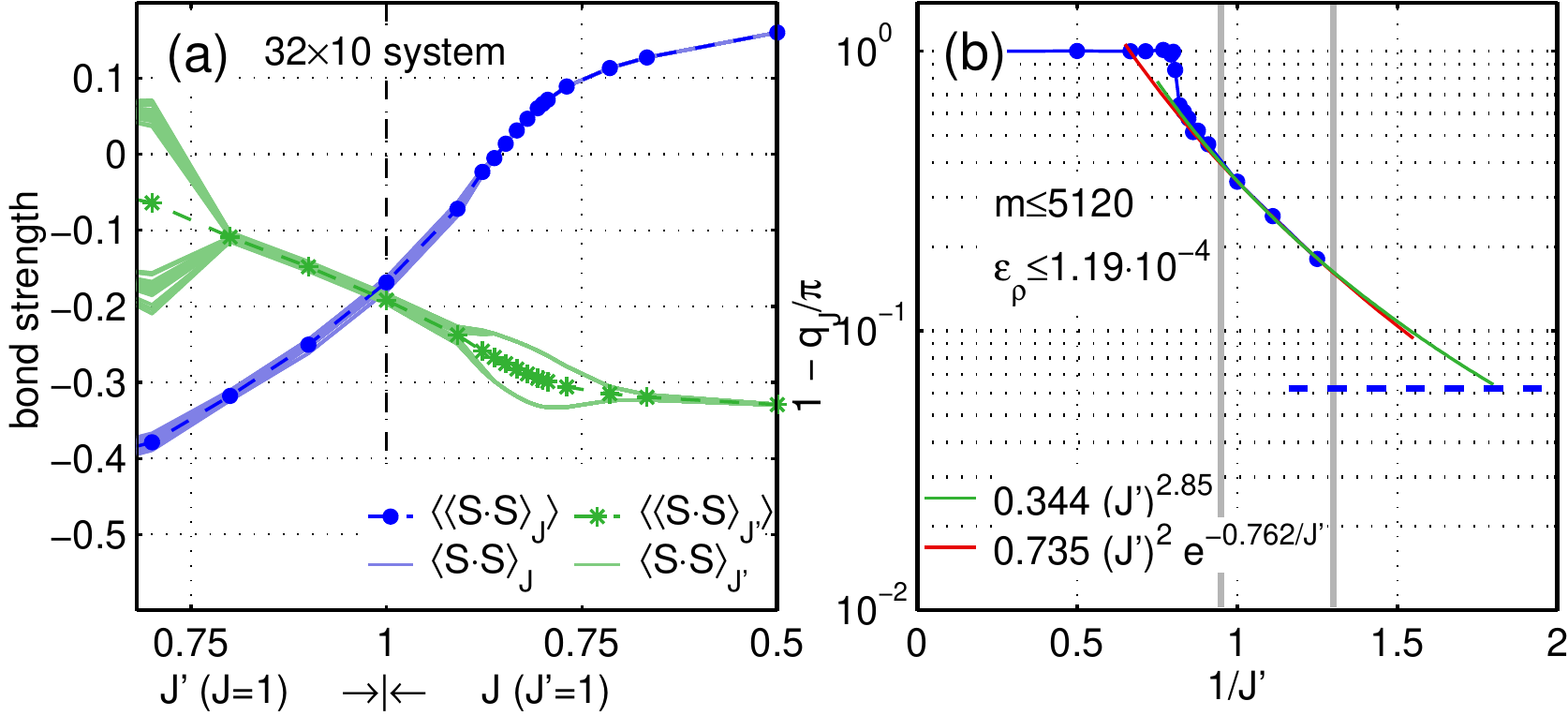}
\end{center}
\caption{(Color online)
Analysis of \SS correlations for width-10 system using \cpsBC with a
similar analysis as in panels (a) and (d) of \Fig{fig:all_128x1} [for
a more detailed description, see caption there]. The system again
shows spontaneous symmetry breaking in terms of a dimerization bubble
for larger \Jp. The regime $\Jp \lesssim 0.8$ suffers strong
numerical limitations [panel (a)]. Panel (b) analyzes the
incommensurate behavior with tentative fits to the regime $\Jp<1$
using $c_3=2$ [\cf \Eq{eq:qJexpfit} and other symmetry-broken systems].
The incommensurate phase terminates at $\Jp \lesssim \Jpc \simeq 1.27$.
The horizontal dashed line again indicates $2/L$, \ie the smallest
$\qJrel/\pi \equiv1-\qJ/\pi$ reachable for given system size. There,
however, the block-entropy has already grown to such an extent that
this limit is no longer reachable reliably numerically.
\vspace{-0.25in} } \label{fig:all_64x5}%
\end{figure}

The same analysis as for the width-2 system in \Fig{fig:all_128x1} is
performed for systems of width-4 [\Fig{fig:all_128x2}], width-6
[\Fig{fig:all_128x3}], width-8 [\Fig{fig:all_64x4}], and width-10
[\Fig{fig:all_64x5}]. All systems analyzed exhibit smoothly changing
incommensurate behavior for finite $\Jp<\Jpc$ with $\Jpc\gtrsim
1.25$. The width-4 system in \Fig{fig:all_128x2} includes reference
data [black pluses in panels (c-d)], with the pinning altered from an
AF-pinning at the left boundary [\cf \Fig{fig:SZanal}] to pinning of
a \emph{single} site [\cf \Fig{fig:SZanal:LH1}]. The data is clearly
consistent with each other, which emphasizes the insensitivity to the
exact details of the pinning at the open boundary and supports a
clear two-chain periodicity normal to the chain direction in the
center of the system.

For comparison, also the spin-gap $\Delta_S$ was calculated for the
systems up to width-8 with rudimentary finite-size scaling only.
\cite{Aristov10} The spin-gap $\Delta_S$ was obtained by calculating
the ground state energy $E_0^S$ for increasing total spin $S$ of a
system with plain cylindrical boundary conditions, \ie in the absence
of pinning fields or smoothing of the boundary. In avoiding fully
periodic boundary conditions for numerical but also physical reasons
[\ie accounting for incommensurate behavior], the open boundary at
the end of the cylinder can carry spinful edge
excitations.\cite{Aristov10} Since these edge states quickly decouple
with increasing system length, they can and do lie within the
spin-gap for the width-$4,6,8,\ldots$ systems. Thus the total spin
$S$ was increased until a true bulk excitation was observed in the
data, \ie the measured $\Szx$ data was no longer exponentially
confined to the boundary. The energy of this state relative to the
global ground state was used to estimate the spin-gap $\Delta_S$.
\cite{Aristov10}

\begin{figure}[tb!]
\begin{center}
\includegraphics[width=0.90\linewidth]{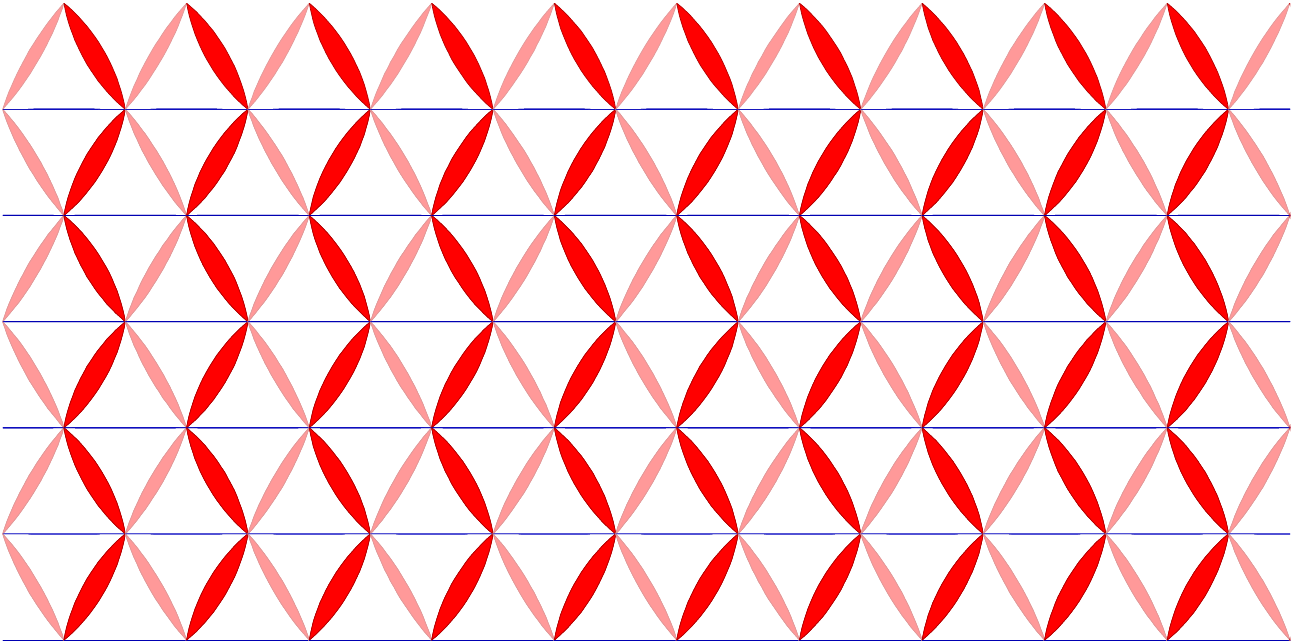}
\end{center}
\caption{(Color online)
Spontaneously symmetry-broken ground state of the $64\times6$ system
[\Fig{fig:all_128x3} at $\Jp=1.16$, having $m=4096$ states kept,
with the chain coupling \Jp chosen such that the intrachain
bond strength \SSJ just crosses zero in \Figp{fig:all_128x3}{a}].
The figure shows the extremely uniform \SSJ and \SSJp across the central
region of the system, having $\SSJp \in \{ -0.3453, -0.2039 \}$
and $\SSJ = 0.0038$, with deviations below given accuracy. This
underlines the \emph{in}-sensitivity to the open boundaries having
\cpsBC. With \SSJ [horizontal bonds] still slightly positive, it is
indicated in blue (black) \vs red (gray) for negative values. The
weaker interchain bond is shown in lighter color for increased
contrast.
} \label{fig:dmz} %
\end{figure}

\subsubsection{Intermediate chain coupling}

The major striking effect seen in the wider systems is the
symmetry-broken alternation of the nearest-neighbor exchange
correlation (to be referred to as dimerization) for intermediate \Jp,
as seen in \Figs{fig:all_128x1}-\ref{fig:all_64x5}. The dimerization
bubble in the \SSJp data, which is strongly visible for width-2
[\Figp{fig:all_128x1}{a}], \emph{disappears} for width-4
[\Figp{fig:all_128x2}{a}] and width-8 [\Figp{fig:all_64x4}{a}], while
it clearly reappears in ever weaker form for width-6
[\Figp{fig:all_128x3}{a}] and width-10 [\Figp{fig:all_64x5}{a}].
While the strength of the dimerization, where present, clearly
weakens for smaller \Jp, it nevertheless appears to persist for
finite $\Jp<1$.

A typical symmetry-broken state for the width-6 system is shown in
\Fig{fig:dmz}, with a similar pattern arising for the width-10
system. Here \Jp was chosen such that the bond strength \SSJ along
the chains just crosses zero [\cf \Figp{fig:all_128x3}{a}]. Note that
a dimerization pattern as in \Fig{fig:dmz} has been recently also
observed on an isotropic four-leg triangular ladder with additional
ring exchanges. \cite{Block11} Overall, the dimerization seen here
suggests a qualitative difference of the systems of width $4n+2$
(symmetry-broken systems), with $n$ an integer, to systems of width
$4n$ (uniform systems), while nevertheless, a two-chain periodicity
perpendicular to the chains is maintained in either case.
Equivalently, this translates into an even-odd effect in the number
of laterally coupled zigzag chains. As the dimerization clearly
weakens with increasing system width, however, in the thermodynamic
limit the dimerization is expected to vanish completely, resulting in
a consistent picture independent of the actual system width.

The reoccurrence of the dimerization in the width ($4n+2$) systems in
Figs.~\ref{fig:all_128x3} and \ref{fig:all_64x5} is also reflected in
several other quantities, similar to what has already been seen in
the width-2 system in \Fig{fig:all_128x1}. Specifically, in the
parameter range where the dimerization is strongest [\eg where the
lower branch in the \SSJp bubble reaches a minimum in panels (a)],
(i) also a minimum is seen in the effective dimension \Deff in panels
(b), while (ii) at the same time the incommensurate behavior
terminates in panels (c) [panel (b) of \Fig{fig:all_64x5}]. For the
width-2 system [\Fig{fig:all_128x1}], this exactly corresponds to the
Majumdar-Ghosh point, $\Jp=2$, while for the width-6 system
[\Fig{fig:all_128x3}] as well as for the width-10 system
[\Fig{fig:all_64x5}] this occurs at $\Jpc\simeq 1.27$. Interestingly,
in all symmetry-broken cases the strongest dimerization always occurs
around the zero-transition of the bond strength \SSJ along the chains
(see panels a).

In contrast, the non-symmetry-broken width $4n$ systems show an
effectively flat \Deff for $\Jp>1$, as seen for width-4 in
\Figp{fig:all_128x2}{b} and width-8 in \Figp{fig:all_64x4}{b}. At
closer inspection, nevertheless a shallow minimum in \Deff is
discernible, which within the accuracy of our data again also
coincides with the point where the incommensurate behavior
terminates. In contrast to the symmetry-broken systems, this
typically occurs at a somewhat larger \Jp still, \ie at $\Jpc\simeq
1.78$ for the width-4 system [\Figp{fig:all_128x2}{c}], and
$\Jpc\simeq 1.56$ for the width-8 system [\Figp{fig:all_64x4}{c}].
The larger \Jpc is also reflected in a qualitatively different shape
of the curve of the incommensurate wave vector \qJ [panels (c)] as
compared to the symmetry-broken systems.

For the phase boundary where the incommensurate behavior vanishes, a
numerical analysis suggests that $\qJ(\Jp) \sim |\Jp-\Jpc|^{1/2}$ for
$\Jp\to(\Jpc)^-$. This is particularly so for the width-2 system,
while for larger widths the incommensurate data is not as reliable to
make a definitive statement. The reason being, that at the point
where the incommensurate behavior vanishes, typically also the
correlation length $\xi$ becomes shortest, \eg even vanishing for the
width-2 system. Consequently, only a very short spatial range is
accessible to determine \qJ from the $S_z$ data, which for all
systems is much shorter than the actual chain length analyzed. While
the extraction still works relatively well for width-2 and width-4
systems, the \qJ data becomes more noisy for the width-6 system, as
seen, for example, in \Figp{fig:all_128x3}{c} around $\Jpc \simeq
1.27$.

Similarly, also the spin-gap $\Delta_S/\Jp$ reflects the
qualitatively different behavior of the non-symmetry-broken width
$4n$ systems [\Figp{fig:all_128x2}{c} and \Figp{fig:all_64x4}{c}], in
that it saturates for large $\Jp$ at a finite value. This value,
however, appears to diminish rapidly with increasing width. For the
symmetry-broken systems of width-2 and width-6, on the other hand,
the spin-gap vanishes for large \Jp. Both sets of systems lead us to
conclude that the spin-gap vanishes in the thermodynamic limit.

\subsubsection{Small chain couplings}

The small \Jp regime is increasingly affected by finite size effects
and limited numerical resources for the wider systems, where the
entanglement across the chains increases strongly. This limits the
numerically accessible range. For the width-4 system in
\Fig{fig:all_128x2}, for $\Jp\lesssim 0.5$ a slight spread is seen in
the individual bond correlations [solid lines] in panel (a), and more
pronouncedly, in panel (b) where the effective dimension \Deff is
cutoff by the maximum number of states that could be kept [$m\le
7000$]. Similar to the width-2 system, \Deff shows a strong
exponential increase for intermediate decreasing $\Jp<\Jpc\simeq
1.78$. For the width-6 system, strong convergence issues arise
for $\Jp\lesssim 0.6$ [\Figp{fig:all_128x3}{a}], for the width-8
system for $\Jp\lesssim 0.7$ [\Figp{fig:all_64x4}{a}], and for the
width-10 system for $\Jp\lesssim 0.8$ [\Figp{fig:all_64x5}{a}]. In
the latter case, the accuracy is already also compromised for
intermediate \Jp, as seen by the slight spread in the individual bond
data for $\Jp\lesssim1.10$.

Bearing in mind this limited numerical accessibility of small \Jp,
the incommensurate behavior for smaller \Jp is analyzed exactly the
same way as for the width-2 system in \Figp{fig:all_128x1}{d} for the
width-4 [\Figp{fig:all_128x2}{d}], width-6 [\Figp{fig:all_128x3}{d}],
width-8 [\Figp{fig:all_64x4}{d}], and the width-10 system
[\Figp{fig:all_64x5}{b}]. The data was fitted both, with an
exponential fit as in \Eq{eq:qJexpfit}, as well as with a plain
polynomial fit. Interestingly, for all systems from width-2 to
width-10, the plain polynomial fit $\qJ \sim (\Jp)^3$ does represent
a very close fit, in agreement with [\onlinecite{Ghamari11}].
However, similar to the discussion of the width-2 system, there
appear systematic deviations which can be improved upon by using an
exponential fit. This is clearly seen for the width-4 system [see
inset to \Figp{fig:all_128x2}{d}], and to a somewhat lesser degree
given numerical limitations for the width-6 [\Figp{fig:all_128x3}{d}]
or width-8 system [\Figp{fig:all_64x4}{d}].

\begin{figure}[tbp!]
\begin{center}
\includegraphics[width=0.9\linewidth]{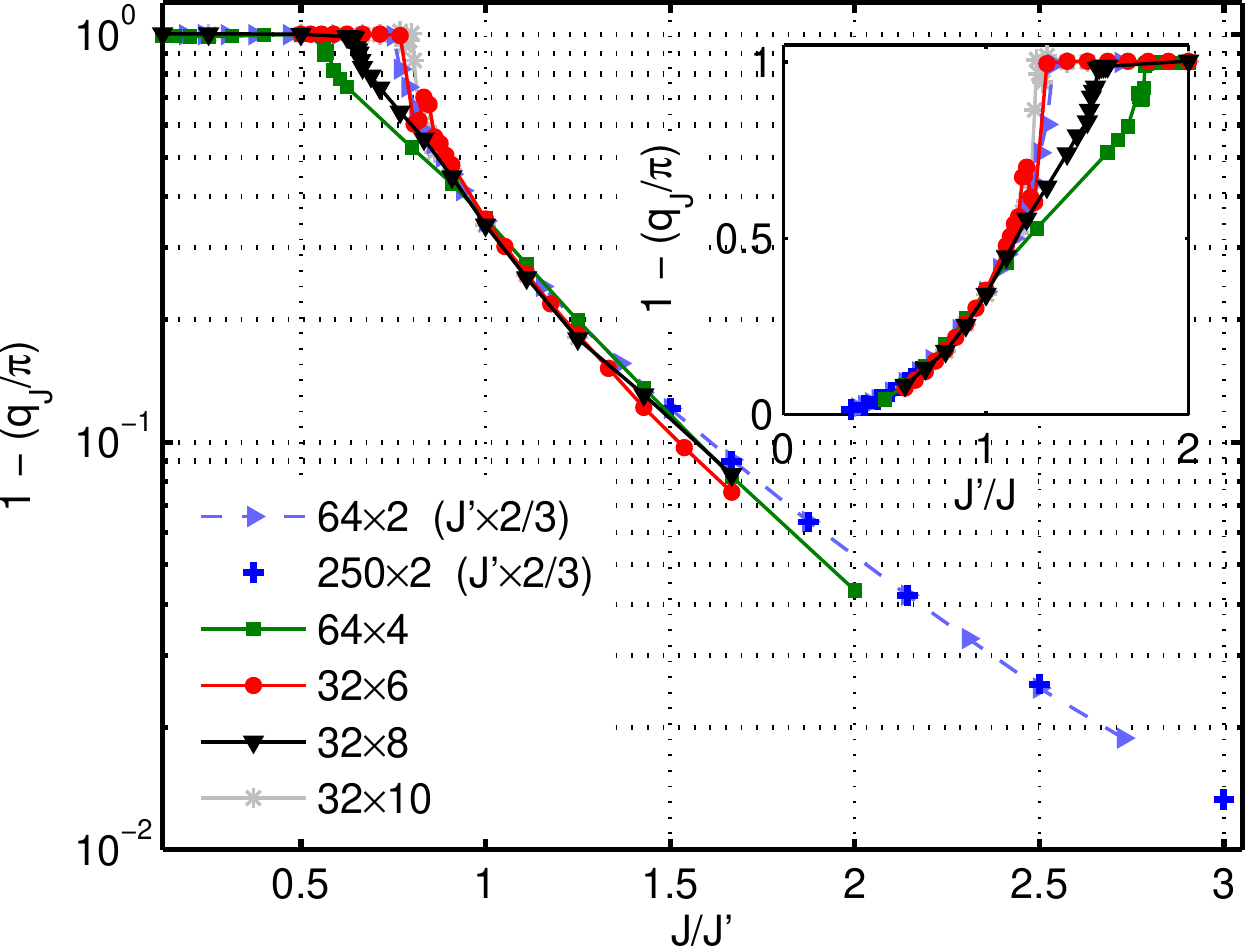}
\end{center}
\caption{
(Color online)
Summarized incommensurate data $\qJrel \equiv \pi - \qJ$ for width-2
to width-10 systems [Figs.~\ref{fig:all_128x1}-\ref{fig:all_64x4},
panel (d), and \Figp{fig:all_64x5}{b}, respectively]. For wider
systems, the incommensurate phase terminates at $\Jpc\simeq 1.25$.
The inset shows the same data vs. \Jp on a linear scale.
} \label{fig:q-all}%
\end{figure}

Finally, the incommensurate data of all systems analyzed
[Figs.~\ref{fig:all_128x1}-\ref{fig:all_64x4}(d) and
\Figp{fig:all_64x5}{b}] is summarized in \Fig{fig:q-all}. Since the
data for the width-2 system is calculated without periodic wrapping
(as this just doubles the strength of the interactions of
\emph{existing} bonds between the chains), a factor of $2/3$ was
applied onto \Jp for the width-2 system such that the incommensurate
data visibly coincides at $\Jp=1$ with the data from the wider
systems.
With this, for smaller \Jp (large $J/\Jp$), the incommensurate data
shows little qualitative and quantitative differences. This supports
the intuitive notion that as the chains become more and more
independent, the dependence of the incommensurate behavior on the
actual system width also weakens. In particular, none of the data
indicates that the incommensurability may vanish for small but finite
\Jp.

\begin{figure}[tbp!]
\begin{center}
\includegraphics[width=0.9\linewidth]{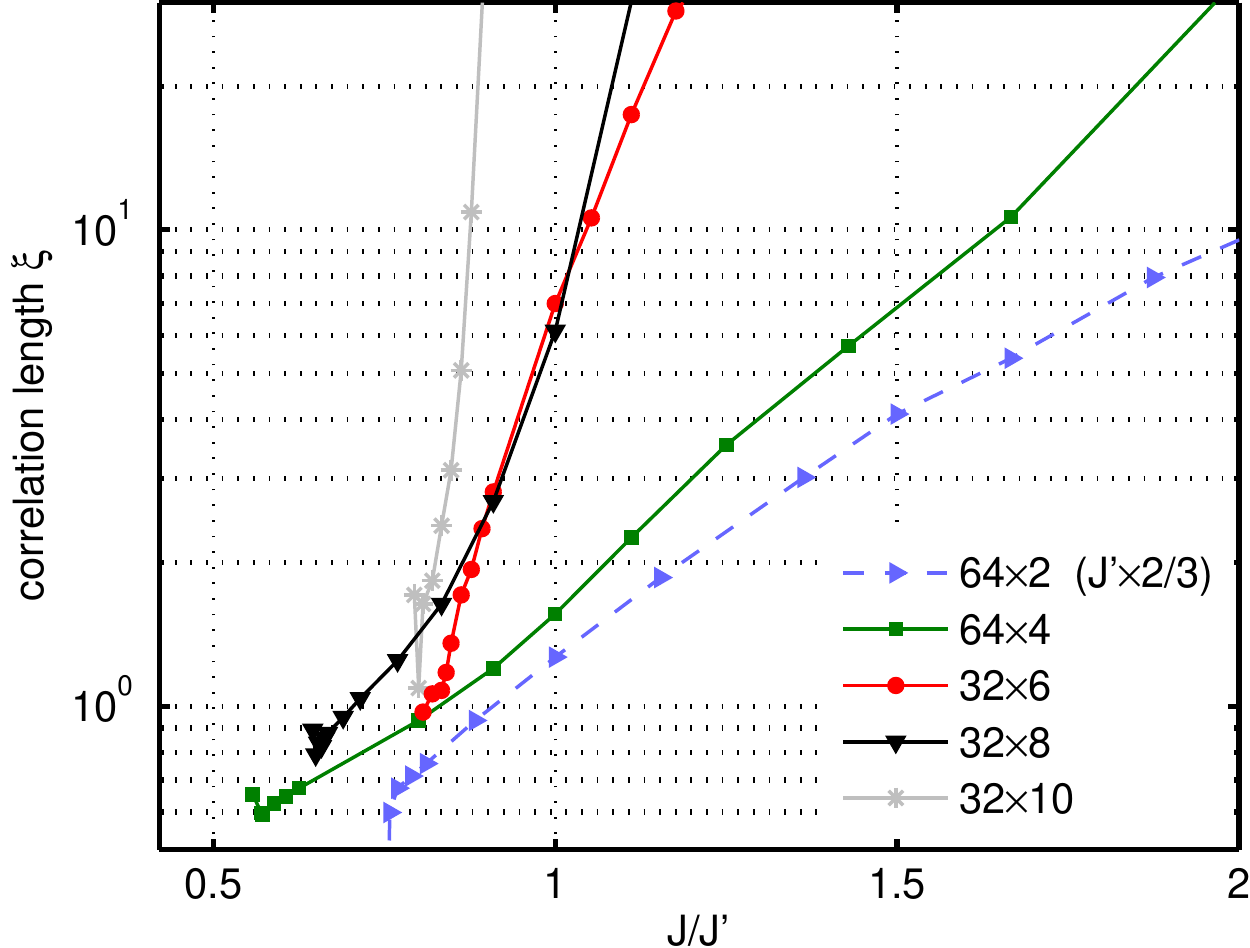}
\end{center}
\caption{ (Color online) Combined data of correlation length defined
through \Eq{eq:CosSin} for width-2 to width-10 systems. The
correlation length is shown only in the parameter regime where the
systems show incommensurate behavior. Outside this range [\ie for
large $\Jp/J$], depending on system width, exponential decay can
be replaced by algebraic decay.
} \label{fig:xi-all}%
\end{figure}

\subsubsection{Correlation length}

In contrast to the incommensurate wave vectors, the correlation
length $\xi$ still shows a pronounced dependence on the system width.
Following the analysis in \Eq{eq:CosSin}, in the incommensurate
regime aside from the oscillating behavior that determines \qJ, a
clear exponential decay is observed and fitted in the central area of
the system away from the open left and right boundaries [\cf
Figs.~\ref{fig:SZanal} and \ref{fig:SZanal:LH1}]. The combined
results for width 2 to 10 are shown in \Fig{fig:xi-all}. The
horizontal axis of the width-2 system again has been scaled the same
way as shown and discussed with \Fig{fig:q-all}. Considering the
qualitative difference between width $4n$ and width ($4n+2)$ systems
then, only width-$(2,6,10,\ldots)$ or width-$(4,8,\ldots)$ may be
directly comparable. This strongly limits finite-size analysis in
terms of the system width. While the correlation length strongly
grows with the width of the systems, consistent with the fact, for
example, that the isotropic case has finite magnetization,
\cite{White07} nevertheless, finite-size scaling in the width of the
system would be crucial in the explicit determination of the
existence of magnetization for arbitrary \Jp in the thermodynamic
limit. This is thus beyond the scope of the present paper.

\section{Summary and Outlook}

The incommensurate correlations on the anisotropic spin-half
Heisenberg lattice have been analyzed over a wide range of chain
couplings $J'/J$. The incommensurate behavior in terms of the
Brillouin zone of a single chain is found to change smoothly from
$\qJ\to\pi$ for weak chain coupling to $\qJ=0$ for $\Jp \ge \Jpc>1$.
In particular, our results are consistent with the $120^\circ$ order
for the isotropic lattice, which is also reflected in the crossing of
\qJ with the classical incommensurability \qJclass at $\Jp=1$ in
\Figs{fig:all_128x2}-\ref{fig:all_64x4}(c). Away from the isotropic
point, the $120^\circ$ order in the spin correlations changes
\emph{smoothly} into the 1D-AF correlations for $\Jp<1$ or into the
square AF correlations for $\Jp\ge\Jpc$. Note that the emphasis here
is on the relative order of \emph{spin correlations}, rather than
explicit magnetization. \cite{White07} The latter is out of the scope
of this paper and thus left as an outlook.

Given the strong frustration in the system, one may expect that for
smaller interchain couplings $\Jp$ the actual data becomes less
sensitive to the width of the system \cite{Ghamari11} [see
\Fig{fig:q-all}]. Therefore already the narrower even-width systems
provide a good qualitative description of the two-dimensional
triangular lattice in the regime of small $\Jp$. Finite size effects
on our cylinders include symmetry-broken and non-symmetry-broken
ground states as for width $(4n+2)$ and $4n$ systems, respectively,
so extrapolations in the width should separate these two
classes.\cite{Ghamari11,Starykh07}
From the analysis of the incommensurate data, we find that
exponential fits of the form \Eq{eq:qJexpfit} fit the data for the
incommensurate wave vectors best. While the accessible range is
limited to finite \Jp, we nevertheless see very systematic behavior
for smaller \Jp down to $\Jp\gtrsim 0.5$ where the correlations
between the chains are already strongly reduced due to inherent
frustration. We take this as evidence that the exponential behavior
is valid down to \Jp=0. That is, the incommensurate behavior remains
present for any finite $0<\Jp<\Jpc$. Given the derived exponential
fits, one may estimate the required system sizes for $\Jp<0.5$.
Taking $\Jp=0.2$ for the width-4 (width-6) system, for example, a
system length of $\gtrsim 8700$ sites ($\gtrsim 3400$ sites) would be
required, respectively. From a DMRG point of view, this is completely
out of reach at this stage. It needs to be seen to what extent
recently emerging infinite size algorithms, such as
iTEBD\cite{Vidal07} or iDMRG \cite{McCulloch07_iDMRG} will be able to
deal with this kind of situation while bearing in mind that
incommensurate correlations with an (exponentially) large underlying
wave length represent a delicate issue.

Interestingly, a very recent quantum simulation in terms of cold
atoms has been performed on the anisotropic triangular lattice
\cite{Struck11} that also suggests that spiral correlation persist
down to $\Jp\to 0$, which is thus consistent with our results.

\begin{acknowledgements}

We want to thank Catherine Kallin for helpful discussions. This work
has received support from the German science foundation (DFG: SFB631,
NIM, and WE4819/1-1) and the NSF under DMR 0907500, and was made
possible by extensive usage of two supercomputing clusters: Greenplanet at UC
Irvine and the Leibnitz Rechenzentrum (LRZ) of the
Bavarian Academy of Sciences.

\end{acknowledgements}


%

\end{document}